\documentclass[journal]{IEEEtranTCOM}

\ifCLASSINFOpdf
\else
\fi

\usepackage{amsmath}
\usepackage[utf8]{inputenc}
\usepackage{cite}
\usepackage{amsmath,amssymb,amsfonts}
\usepackage{algorithmic}
\usepackage{graphicx}
\usepackage{textcomp}
\usepackage{xcolor}
\usepackage{float}
\usepackage{physics}
\usepackage{makecell}

\usepackage{enumerate} 
\usepackage{subfigure}

\def\BibTeX{{\rm B\kern-.05em{\sc i\kern-.025em b}\kern-.08em
    T\kern-.1667em\lower.7ex\hbox{E}\kern-.125emX}}

\begin{document}
%
% paper title
% Titles are generally capitalized except for words such as a, an, and, as,
% at, but, by, for, in, nor, of, on, or, the, to and up, which are usually
% not capitalized unless they are the first or last word of the title.
% Linebreaks \\ can be used within to get better formatting as desired.
% Do not put math or special symbols in the title.
\title{Symbol-Level Noise-Guessing Decoding with Antenna Sorting for URLLC Massive MIMO}
%
%
% author names and IEEE memberships
% note positions of commas and nonbreaking spaces ( ~ ) LaTeX will not break
% a structure at a ~ so this keeps an author's name from being broken across
% two lines.
% use \thanks{} to gain access to the first footnote area
% a separate \thanks must be used for each paragraph as LaTeX2e's \thanks
% was not built to handle multiple paragraphs
%

\author{Sahar Allahkaram,~\IEEEmembership{Student Member,~IEEE,}
        Francisco A. Monteiro,~\IEEEmembership{Member,~IEEE,}
        \\ 
        and~Ioannis Chatzigeorgiou,~\IEEEmembership{Senior Member,~IEEE}% <-this % stops a space
        
\thanks{S.~Allahkaram~is~with~Instituto~de~Telecommunica\c{c}\~{o}es,~and~\mbox{ISCTE-Instituto} Universit\'{a}rio de Lisboa, Portugal, e-mail: sahar.allahkaram@lx.it.pt.}
\thanks{F.~A.~Monteiro~is~with~Instituto~de~Telecommunica\c{c}\~{o}es,~and~\mbox{ISCTE-Instituto} Universit\'{a}rio de Lisboa, Portugal, e-mail: francisco.monteiro@lx.it.pt.}
\thanks{I. Chatzigeorgiou is with the School of \mbox{Computing \& Communications}, Lancaster University, UK, e-mail: i.chatzigeorgiou@lancaster.ac.uk.}
\thanks{Some results in this paper have been previously presented at the IEEE $96^{th}$ Vehicular Technology Conference (VTC2022-Fall), 2022.}
}

% make the title area
\maketitle

% As a general rule, do not put math, special symbols or citations
% in the abstract or keywords.
\begin{abstract}
Supporting ultra-reliable and low-latency communication (URLLC) is a challenge in current wireless systems. Channel codes that generate large codewords improve reliability but necessitate the use of interleavers, which introduce undesirable latency. Only short codewords can eliminate the requirement for interleaving and reduce decoding latency. This paper suggests a coding and decoding method which, when combined with the high spectral efficiency of spatial multiplexing, can provide URLLC over a fading channel. Random linear coding and high-order modulation are used to transmit information over a massive multiple-input multiple-output (mMIMO) channel, followed by zero-forcing detection and guessing random additive noise decoding (GRAND) at a receiver. A variant of GRAND, called symbol-level GRAND, originally proposed for single-antenna systems that employ high-order modulation schemes, is generalized to spatial multiplexing. The paper studies the impact of the orthogonality defect of the underlying mMIMO lattice on symbol-level GRAND, and proposes to leverage side-information that comes from the mMIMO channel-state information and relates to the reliability of each receive antenna. This induces an antenna sorting step, which further reduces decoding complexity by over 80\% when compared to bit-level GRAND.
\end{abstract}

% Note that keywords are not normally used for peerreview papers.

\IEEEpeerreviewmaketitle

\section{Introduction}
In addition to other crucial requirements for the sixth generation (6G) of wireless networks, such as low energy consumption, high scalability, stability, security, and ubiquitous connectivity, the physical layer of wireless communications will have to significantly contribute to the goal of ultra-reliable and low-latency communications (URLLC). To meet the important requirements of applications like the industrial internet of things (IIoT), virtual reality, or self-driving cars, URLLC's main objectives are to reduce latency to 1 ms while concurrently guaranteeing at least 99.999\% dependability\cite{Nouri2020}. Using error-correcting codes with short codewords is one way of achieving the sought low-latency  objective, because that allows to discard  the interleavers that are typically employed in wireless links to make the errors look independent and identically distributed (i.i.d.) \cite{An2022}. However, developing codes with large codewords was prioritized in pre-5G systems to reach Shannon's capacity \cite{Shannon1948,Gallager1973}. An interest in codes from the 1960s, such as Reed-Solomon and BCH codes, was rekindled, aiming at URLLC applications \cite{vucetic19}. While these codes can have short codewords, they only exist for a limited number of code rates. Contrary to that, random linear codes (RLCs) can be constructed with any code rate, even though decoding long RLCs is impractical \cite{Becker2012}.

It was previously known that short random linear codes (RLCs) could be decoded using trellis decoding \cite{Wolf_1978, ShuLin_TIT_1993, Kschischang_Sorokine_TIT_1995, Forney_TIT_1998, Amir_thesis_1997,Monteiro_Kschischang_11, Honary_book_97,ShuLin_book_98} or information set decoders \cite{Coffey_Goodman_90}, however, given the historical emphasis on capacity-achieving codes (with long codewords), that path of research seems to have been abandoned by the coding community.
Recently, noise-guessing decoding has been proposed as a universal decoding technique for codes with moderate length or sufficiently high rate, which are particularly suited for wireless URLLC \cite{Duffy2019}. The method, known as guessing random additive noise decoding (GRAND) allows maximum likelihood (ML) decoding with a considerably reduced complexity, chiefly because it focuses on ``decoding the noise'' rather than the codewords, by taking advantage of the entropy of the noise being much lower than the entropy of the codewords. The sole requirement is that a code membership test exists to decide whether some word is a valid codeword. Consequently, GRAND can perform ML decoding of binary or nonbinary linear codes without the need to compute a trellis or store a large table.

GRAND opened doors to using RLCs, known to be capacity-achieving in the asymptotic regime (i.e., with infinite length codewords) in the binary symmetric channel (BSC) \cite{Shannon1948, Gallager1973}, and they also reach capacity in the finite-blocklength regime \cite{Duffy2019, Polyanskiy_2010, Polyanskiy_2014}, which is the regime of interest for URLLC applications.
Several recent research works have shown that RLCs supersede the performance of polar codes of the same length and rate in the classical case \cite{vucetic23,allahkaram2022urllc}. Most importantly, while off-the-shelf nonrandom codes, such as polar codes, do not exist for any desired pair of code length and code rate, one has great flexibility of choice regarding the length and code rate when employing RLCs with GRAND \cite{An2022,Duffy2019, Duffy2022}. For higher spectral efficiency, GRAND has been proposed in combination with massive multiple input-multiple-output (mMIMO) in \cite{allahkaram2022urllc}. The ideas behind GRAND have also been adapted to allow the decoding of quantum random linear codes in a practical manner \cite{Cruz2022}, and also to decode quantum stabilizer codes with a given structure (i.e., non-random known codes) \cite{Chandra2023}.
%{Patent_2020a, Patent_2020b, Patent_2021}

Symbol-level GRAND has been recently proposed in \cite{Chatzigeorgiou_Monteiro_2023} for single-input single-output (SISO) block fading channels, of which the additive white noise Gaussian noise (AWGN) channel is a special case. Symbol-level GRAND attains significantly faster decoding than the original bit-level GRAND. In \cite{An2022}, the authors have suggested modifying GRAND to use knowledge about the adopted modulation scheme for channels with memory. Symbol-level GRAND takes a different approach: it relies on a closed-form expression for the probability that the input stream of bits contains a specific combination of bit strings representing various constellation symbols. These constellation symbols have  different numbers of nearest and next-nearest neighbors. When the transmission is done over a block fading channel, the expression allows to order the error patterns according to their likelihood. 

With the aim of attaining the URLLC objectives, this work integrates $M$-ary quadrature phase modulation ($M$-QAM), RLC encoding and symbol-level GRAND into a mMIMO system that employs zero-forcing (ZF) detection. While RLCs cater for the sought-after high reliability and GRAND offers reduced decoding complexity, mMIMO techniques enable high spectral efficiency through spatial multiplexing. We explain that symbol-level GRAND can be directly extended to mMIMO, if strong channel hardening (CH) conditions are assumed. Furthermore, we show that the considered mMIMO system can cope with adverse CH conditions, if the symbols at the output of the ZF detector are ordered according to their reliability, which can be derived from channel state information (CSI).

The optimized re-ordering of symbols can be seen as an antenna sorting problem. Antenna sorting has been known to greatly impact the detection performance of MIMO systems that use a small number of spatial streams. Optimal antenna sorting strategies that rely on the notion of the \textit{effective} signal-to-noise ratio (SNR) of a stream at the output of the MIMO detector \cite{Jiang2011, Monteiro2012} have been devised for different MIMO detection methods. For example, antenna sorting was used to increase the performance of V-BLAST detectors \cite{Monteiro2012}, or to simultaneously improve the performance and reduce the complexity of sphere decoders \cite{1494671}. In this paper, we use the effective SNR after ZF detection as a sorting metric akin to the reliability of the QAM symbols, which carry the bit strings that build up the codewords. The proposed antenna sorting method further reduces the complexity of symbol-level GRAND.

The paper starts by describing RLCs in Section~\ref{sec:RLCs}, then bit-level GRAND is described in Section~\ref{sec:GRAND}. The system model is detailed in Section~\ref{sec:model}. Section~\ref{sec:sorting} shows how symbol reliability can be obtained from CSI and how antenna sorting should be implemented. A detailed analysis of the operation of symbol-level GRAND is presented in Section~\ref{sec:symbolLevelGRAND}. Section~\ref{sec:analysis} derives a lower bound on performance, considering perfect channel hardening (PCH), and then Section~\ref{sec:results} shows performance and complexity results of the proposed scheme. Section~\ref{sec:conclusion} summarizes key conclusions.

% The very first letter is a 2 line initial drop letter followed
% by the rest of the first word in caps.
% 
% form to use if the first word consists of a single letter:
% \IEEEPARstart{A}{demo} file is ....
% 
% form to use if you need the single drop letter followed by
% normal text (unknown if ever used by the IEEE):
% \IEEEPARstart{A}{}demo file is ....
% 
% Some journals put the first two words in caps:
% \IEEEPARstart{T}{his demo} file is ....
% 
% Here we have the typical use of a "T" for an initial drop letter
% and "HIS" in caps to complete the first word.
% You must have at least 2 lines in the paragraph with the drop letter
% (should never be an issue)

\section{Random Linear Codes}
\label{sec:RLCs}

Linear block codes can be concisely represented by generator matrices. The encoding operation of an linear block code can be described by the multiplication of an input information vector of length $k$ with a $k\times n$ generator matrix to obtain an output vector, referred to as a codeword, of length $n>k$. In the case of RLCs, the elements of the generator matrix are selected uniformly and at random from a Galois field $\mathbb{F}_q=\{0,1,\ldots,q-1\}$. The entries of the input and output vectors are also elements of $\mathbb{F}_q$, where $q=2$ in the case of binary codes. The ratio $R=k/n$ is the code rate.

The most common method for decoding linear block codes, including RLCs, is syndrome decoding, which relies on the $(n-k)\times n$ parity-check matrix. The parity-check matrix is designed such that the product of the generator matrix and the transpose of the parity-check matrix is the $k\times (n-k)$ zero matrix. The multiplication of the parity-check matrix with a potentially erroneous received word generates a vector of  $n-k$ bits, known as the syndrome. Syndrome decoding achieves ML decoding, but a lookup table is required for the storage of possible syndromes and respective coset leaders. For example, assume that $q=2$ and suppose that we wish to correct received words that contain up to a threshold of $w_{th}$ bit errors. The number of error patterns that need to be considered is given by $\sum_{t=0}^{w_{th}} \binom{n}{t}$. However, the number of all possible syndromes is $2^{n-k}$. For large values of $w$ and high code rates, that is $R\rightarrow 1$ and thus $n\rightarrow k$, the relationship $\sum_{t=0}^{w_{th}} \binom{n}{t} \gg 2^{(n-k)}$ holds, therefore the error correction capability of the code is limited because a wide variety of error patterns result in the same syndrome. Choosing the coset leader associated with each particular syndrome depends on side information regarding the \textit{a priori} probability of each error pattern. Over AWGN, the chosen coset leaders should be the error patterns with the lowest Hamming weight, which leads to ML decoding.

\section{Guessing random additive noise decoding (GRAND)}
\label{sec:GRAND}

A workable technique for decoding RLCs has recently appeared with the advent of GRAND-based algorithms\cite{An2022},\cite{Duffy2019},\cite{Duffy2020}. GRAND achieves ML decoding by  ``decoding the noise'' that corrupted the codeword rather than attempting to decode the potential codewords\cite{Duffy2019}. GRAND is a universal decoder that can be applied to block codes of moderate length and high code rate, regardless of whether the code is binary or multi-level, random, or has some other kind of mathematical structure (such as polar codes \cite{Duffy2020, Duffy2021}, BCH codes \cite{Abbas, Duffy2021}, or Hamming codes). The only prerequisite for GRAND is a membership test to determine whether a word qualifies as a codeword. 
The test for RLCs is based on the syndrome of the codewords. Most importantly, these RLCs can be built with any desired \textit{codelength} and \textit{rate} which is a huge benefit for fitting any needed code numerology for a particular application.

In comparison with an exhaustive search (i.e. comparing the received codeword with every codeword in the codebook), or even with syndrome-based decoding, guessing the noise becomes substantially faster. This reduction in the search time is due to the low entropy of the noise, which translates into having a manageable list size for the potential error patterns affecting a codeword. A natural consequence of the concept is that any knowledge about noise statistics can be used to introduce search constraints, reduce the search space and speed up the search process \cite{An2022}. In fact, any extra information regarding the \textit{a priori} probability of the error patterns should be used, and that concept is at the core of the proposals in this paper.

When there is access to soft information about a received bit (or symbol), that soft information can serve as a reliability metric for the bit (or symbol) \cite{Duffy2021, solomon_soft_2020}. Noise guessing should prioritize the least reliable positions in order to increase the probability of finding the correct noise pattern and reduce the decoding time. For practical reasons, it is preferable to first sort the bits or symbols in increasing reliability order and then modify the bits (or symbols) in a ``natural counting order'' when guessing the error patterns.
This mechanism can be used by default, while the sorting is delegated to a preprocessing unit.
 
\begin{figure}[t]
\centering
  \includegraphics[width=1\columnwidth, clip=true, draft=false]{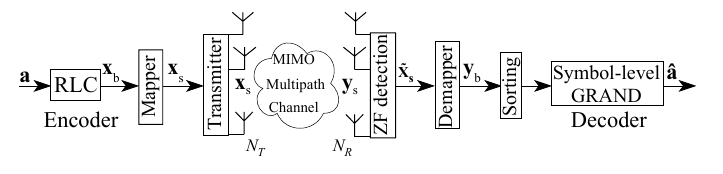}
    \caption{System model for coded mMIMO URLLC.} 
    \label{fig:system}
\end{figure}

\section{System model for coded massive MIMO}
\label{sec:model}

A coded massive MIMO system is considered, making use of an RLC encoder at the transmitter and symbol-level GRAND at the receiver. A block of $k$ information bits is mapped onto a $n$-bit codeword and sent ``over the air'' via spatial multiplexing, as illustrated in Fig. \ref{fig:system}. This process may be repeated when transmitting longer information streams by dividing the bit stream into blocks of size $k$.

\subsection{RLC encoding and spatial multiplexing with mMIMO}
\label{sec:mMIMO}

A block $\textbf{a}$ of $k$ i.i.d. information bits is linearly encoded into a codeword $\mathbf{x}_\text{b}$ of length $n$ using a systematic binary RLC with rate $R=k/n$, denoted by RLC $(n,k)$. The RLC $(n,k)$ defines a codebook $\mathcal{C}$ with $2^k=2^{nR}$ codewords of length~$n$, which constitutes a linear subspace of the discrete vector space $\mathbb{F}_{2}^{n}$. Although the minimum Hamming distance between two codewords impacts the error correction capability of linear block codes, the minimum Hamming distance in RLCs is not as relevant in determining the code's performance \cite[Ch.13]{MacKay2003}. The RLC $(n,k)$ is described by a \textit{random} binary generator matrix $\mathbf{M} \in \mathbb{F}_{2}^{k \times n}$, which acts as the basis matrix for the code subspace, such that $\mathcal{C}=\left\{\textbf{x}_\text{b}  = \textbf{a} \mathbf{M}: \textbf{a} \in \mathbb{F}_{2}^{k}\right\}$. The generator matrix is of the form $\mathbf{M}=\left[\ \mathbf{I}_k \ | \ \mathbf{P} \ \right]$, where $\mathbf{P} \in \mathbb{F}_{2}^{k \times (n-k)}$ is a random binary matrix, and $\mathbf{I}_k$ is the $k \times k$ identity matrix responsible for the systematic part of the encoding.

The $n$ bits, $b_1,\ldots,b_n$, of a codeword are input to a \mbox{$M$-QAM} mapper. The mapper divides the sequence of $n$ bits into $L$ strings of $\log_2(M)$ bits, that is, $L=n/\log_2(M)$, and maps the $L$ strings onto $L$ complex-valued symbols, $s_1,\ldots,s_L$, taken from the alphabet $\mathcal{A} \in \mathbb{C}$ of the $M$-QAM constellation. The cardinality of $\mathcal{A}$ is $\lvert \mathcal{A}\rvert = M$. We denote the $n$-bit codeword and the sequence of $L$ modulated symbols by $\mathbf{x}_\text{b}=[b_1, \ldots, b_n]$ and $\mathbf{x}_\text{s}=[s_1, \ldots, s_L]^T$, respectively. Furthermore, we denote by $\mathcal{S}(s_i)$ the string of $\log_2(M)$ bits that has been mapped onto symbol $s_i$. Thus, the codeword $\mathbf{x}_\text{b}$ can also be written as $\mathbf{x}_\text{b}=[\mathcal{S}{(s_1)}, \ldots, \mathcal{S}{(s_L)}]$. If $E_b$ represents the energy per information bit, then $(k/n)E_b$ is the energy per codeword bit, and $\log_2(M) (k/n)E_b$ is the energy per string of $\log_2(M)$ bits, which also corresponds to the energy per symbol. 

The system can be designed to allow the transmission of $N_c$ codewords in each MIMO channel use. This implies that, when a specific cardinality $M$ is employed for the modulation, the number of transmit antennas is $N_T=N_cL$. Without loss of generality, and to keep the notation simple, we will describe the system for $N_c=1$, where one MIMO burst transmitted from the $N_T$ antennas contains one codeword only (i.e., $N_T=L$). Later, in Section \ref{subsec:sorting}, the generalization for $N_c>1$ will be commented on. A system with $N_c<1$ can also be made operational by adding buffers both at the transmitter and at the receiver, hence creating a full separation between the mMIMO physical layer and channel coding and decoding such that symbol-level GRAND only starts decoding when the $L$ symbols corresponding to a codeword have been received.

The coded signal $\mathbf{x}_\text{s}$ is transmitted over a MIMO Rayleigh fading channel, characterized by the matrix $\mathbf{H} \in \mathbb{C}^{N_T \times N_R}$, where $N_R \gg N_T$ is the number of antennas fitted at the receiver. The received signal $\mathbf{y}_\text{s}= [y_1, ..., y_{N_R} ]^T$ is given by:
\begin{equation}
	\mathbf{y}_\text{s} = \sqrt{\frac{\mathrm{snr}}{N_T}} \mathbf{H}\mathbf{x}_\text{s} + \mathbf{n},
     \label{eq:y_s}
\end{equation}
where
\begin{equation}
    \mathrm{snr} \triangleq \log_2(M) \left(k /n \right) \left(E_b/N_0\right)
    \label{eq:ergodic_SNR}
\end{equation}
is the ergodic SNR at the receiver and $\mathbf{n}= [n_1, ... , n_{N_R} ]^T$ represents the additive noise at the receiver. The entries in both $\mathbf{H}$ and in $\mathbf{n}$ are i.i.d. random variables taken from a complex normal distribution. The entries in $\mathbf{H}$ are taken from $\mathcal{CN}(0,1)$ and the ones in $\mathbf{n}$ are taken from $\mathcal{CN}(0,\sigma_\text{n}^2)$ , with  $\sigma_\text{n}^2=1$. The symbols in $\mathcal{A}$ are normalized to unit average energy, so that $\mathbb{E}\{|s_i|^2\}=1$. The $N_T \times N_R$ matrix $\mathbf{H}$ remains constant during the transmission of $\mathbf{x}_\text{s}$ but changes independently from channel use to channel use. 

\subsection{Zero-forcing detection and symbol-level GRAND}
\label{sec:ZFandGRAND}

ZF detection amounts to applying the Moore-Penrose pseudo-inverse \cite{Lozano18, Monteiro2012}
\begin{equation}
    \mathbf{H^+}=\left(\mathbf{H}^H\mathbf{H}\right)^{-1} \mathbf{H}^H.
    \label{eq:Moore-Penrose}
\end{equation}
The application of \eqref{eq:Moore-Penrose} to \eqref{eq:y_s} at the receiver results in
\begin{equation}  \mathbf{H^+}\mathbf{y}_\text{s}=\sqrt{\frac{\mathrm{snr}}{N_T}} \mathbf{I}_{N_T} \mathbf{x}_\text{s}+\underbrace{\mathbf{H^+}\mathbf{n}}_{\mathbf{u}},\quad
    \label{eq:ZF}
\end{equation}
where $\mathbf{I}_{N_T}$ is the $N_T\times N_T$ identity matrix, and $\mathbf{u} \in \mathbb{C}^{N_T}$ denotes the new (now correlated) noise vector after ZF filtering. Although the performance of ZF detection is rather poor in symmetric MIMO, where $N_R=N_T$, it attains quasi-optimal performance in highly asymmetric MIMO, for example, when $N_R>>N_T$ in an uplink scenario \cite{Lozano18}. In this scenario, which is considered in our system model, the instantaneous SNR for each channel realization approaches its ergodic value at each received data stream after ZF detection. At the same time, the large value of $N_R$ boosts the receiver array gain.
 
After the linear filtering in \eqref{eq:ZF}, a quantization operation $\mathcal{Q}(\cdot)$ is made to the $M$-QAM constellation to obtain the sequence of detected symbols $\tilde{\mathbf{x}}_\text{s}=\mathcal{Q}(\mathbf{H^+}\mathbf{y}_s)=[\tilde{s}_1, \dots, \tilde{s}_L]^T$, which is an estimate of $\mathbf{x}_\text{s}$ corrupted by noise. The detected symbols $\tilde{s}_1, \dots, \tilde{s}_L$ are demapped to bit strings $\mathcal{S}(\tilde{s}_1), \ldots, \mathcal{S}(\tilde{s}_L)$ and reconstruct a word of $n$ bits, denoted by $\mathbf{y}_\text{b}$. The relationship between the reconstructed word $\mathbf{y}_\text{b}$ at the receiver and the codeword $\mathbf{x}_\text{b}$ at the transmitter is $\mathbf{y}_\text{b}=\mathbf{x}_\text{b} \oplus \mathbf{e}_\text{b}$, where $\mathbf{e}_\text{b}$ is the error pattern that has corrupted the transmitted codeword. The operation $\oplus$ denotes modulo-$2$ addition. The word $\mathbf{y}_\text{b}$ is input to symbol-level GRAND, which attempts to estimate $\mathbf{e}_\text{b}$ and infer $\mathbf{x}_\text{b}$ using $\mathbf{\hat{x}}_\text{b}=\mathbf{y}_\text{b} \oplus \mathbf{\hat{e}}_\text{b}$, where $\mathbf{\hat{x}}_\text{b}$ and $\mathbf{\hat{e}}_\text{b}$ are estimates of $\mathbf{x}_\text{b}$ and $\mathbf{e}_\text{b}$, respectively. The first $k$ of the $n$ bits of the estimated codeword $\mathbf{\hat{x}}_\text{b}$ form the block of decoded information bits $\mathbf{\hat{a}}$, as shown in Fig.~\ref{fig:system}.

\section{Symbol reliability and antenna sorting}
\label{sec:sorting}

\subsection{Effective post-processing SNR}
After ZF inversion, the decisions made by the quantizer $\mathcal{Q(\cdot)}$ to obtain $\mathbf{\hat{x}_\text{s}}$ are perturbed by the modified noise vector $\mathbf{u}$ that appears in \eqref{eq:ZF}.
One can show that the output SNR after ZF detection of the $N_T$ incoming signals streams depends on the instantaneous channel realization $\mathbf{H}$ in the following manner \cite{Jiang2011} \cite[sec. 3.1.3]{Monteiro2012}:

\begin{equation}
\begin{aligned}
  \mathrm{snr}_i^{(ZF)}
  & =\frac{\mathrm{snr}}{\left[\left(\mathbf{H}^{H} \mathbf{H}\right)^{-1}\right]_{ii}}\\
  & = \frac{1}{\left[ \mathbf{G}^{-1}\right]_{ii}}\mathrm{snr}=
   g_i \ \mathrm{snr}, \quad 1 \leq i \leq N_T
   \label{SNR-gain}
\end{aligned}
\end{equation}

\noindent where the $g_i$ are defined as the inverses of the diagonal of $\mathbf{G}^{-1}$, for $i=1,\dots,N_T$. Note that $\mathbf{G}$ is the Gram matrix of the lattice spanned by the columns of $\mathbf{H}$ (e.g., \cite{Monteiro_2010}). 
A different definition of $\mathrm{snr}$ is used in \cite{Jiang2011} but that does not change the relation in \eqref{SNR-gain}. This expression provides soft information about the reliability of each symbol, given the one-to-one relation with each antenna stream. This information will be central to sorting the received symbols so that symbol-level GRAND can perform its guesswork of the transmitted symbols starting from the least reliable symbol to the most reliable one.

The value of each $g_i$ in \eqref{SNR-gain} should be as large as possible. In the case of a diagonal $\mathbf{G}$, that maximization happens for a $\mathbf{G}$ with large diagonal elements. If the energy spills over the diagonal, the elements in the diagonal get smaller due to energy conservation arguments. This corresponds to having the off-diagonal elements of $\mathbf{G}$ no longer close to zero due to non-orthogonality of the column vectors of $\mathbf{H}$.

\subsection{Lattice geometry with a finite number of antennas}
The geometry of ZF detection fully determines its detection performance. For ZF to approach ML detection using the Voronoi regions of the underlying the real MIMO lattice, it is necessary that the so-called ZF detection region matches the Voronoi region with a low discrepancy (e.g., \cite{Monteiro2012}). This is the fundamental cause for ZF detection becoming optimal as $N_R$ increases. When $N_R \rightarrow \infty$ the lattice spanned by the columns of $\mathbf{H}$ would be a perfectly orthogonal lattice, and ZF would be optimal. Analytically, this effect can be captured by measuring the  effect of the effective noise $\mathbf{u}$ in (\ref{eq:ZF}). The effect of the ZF filter on that noise power can be tracked by considering the autocorrelation matrix of the new noise $\mathbf{u}= \mathbf{H}^+\mathbf{n}$, calculated as:
\begin{equation}\label{eq:R_u}
\begin{aligned}
\mathbf{R_u} & =\mathbb{E} \left\{ \mathbf{u} \mathbf{u}^H \right\}
    = \mathbb{E} \left\{ \left(\mathbf{H}^+\mathbf{n} \right) \left(\mathbf{H}^+\mathbf{n} \right)^H \right\}
    \\
    & = \mathbb{E} \left\{ \left(\mathbf{H}^+\mathbf{n} \right)
    \left(  \mathbf{n}^H (\mathbf{H}^+)^H  \right) \right\}\\
    & =  \mathbf{H}^+  \mathbb{E} \left\{  \mathbf{n} \mathbf{n}^H \right\} (\mathbf{H}^+)^H = \sigma_\text{n}^2 \mathbf{H}^+ (\mathbf{H}^+)^H,
\end{aligned}
\end{equation}
where the autocorrelation of the original Gaussian noise, $\mathbb{E}  \left\{  \mathbf{n} \mathbf{n}^H \right\} = \mathbf{R_n} = \sigma_\text{n}^2 \mathbf{I}_{N_R}$, has been used. Replacing the Moore-Penrose pseudo-inverse from \eqref{eq:Moore-Penrose} in \eqref{eq:R_u}, and using the definition of the Gram matrix, it is possible to obtain
\begin{equation}\label{eq:correlation}
    \mathbf{R_u} = \sigma_\text{n}^2 \left(\mathbf{H}^H\mathbf{H}\right)^{-1}=\sigma_\text{n}^2 \mathbf{G}^{-1}.
\end{equation}
From both \eqref{SNR-gain} and \eqref{eq:correlation}, one can see that ZF detection always causes noise enhancement in the case of real-world channels (with a finite $N_R$).

The noise amplification of ZF detection can be geometrically interpreted using lattices. Let $\mathcal{G}=\mathbb{Z}+i\mathbb{Z}$ denote the set of Gaussian integers. A complex lattice is defined as $\Lambda=\{ \mathbf{H} \mathbf{z}: \mathbf{z} \in \mathcal{G}^{N_T \times 1} \}$. For a lattice basis $\mathbf{H} \in \mathbb{C}^{N_R \times N_T}$, the lattice has rank $N_T$, and lives in a $N_R$-dimensional space. The volume of the fundamental region of the lattice is $\operatorname{vol}(\Lambda)=\sqrt{\operatorname{det}\left(\mathbf{H}^H \mathbf{H}\right)}=\sqrt{\operatorname{det}(\mathbf{G})}$. In the case of square matrices, this simplifies to $\operatorname{vol}(\Lambda)=\operatorname{det}(\mathbf{H})$. 
In MIMO detection it is preferable to use the real-valued equivalent lattice, defined as $\Lambda_\mathbb{R} =\{ \mathcal{H} \mathbf{z}: \mathbf{z} \in \mathbb{Z}^{2N_T \times 1} \}$, having rank $2N_T$, and living in $2N_R$ dimensions. It uses the equivalent \textit{real-valued} basis $\mathcal{H} \in \mathbb{R}^{2N_R \times 2N_T}$, constructed from the complex basis $\mathbf{H}$ \cite{Monteiro2012}.

Noise amplification is larger when there is a large mismatch between the so-called \textit{ZF detection region} and the \textit{Voronoi region} of that lattice. Given that the first is always a $2N_T$-dimensional parallelotope, that match can only be perfect in the case of a perfectly orthogonal lattice.
To measure how orthogonal a lattice $\Lambda_\mathbb{R}$ is, one can use the so-called \textit{orthogonality defect} (OD), a metric originally proposed to analyze the detection of conventional MIMO \cite{LLL}.
The OD of a lattice spanned by a real basis $\mathcal{H}$ is defined as:
\begin{equation}
    od(\mathcal{H})=\frac{\prod_{i=1}^{2N_T}\left\|\mathbf{h}_i\right\|}{\operatorname{vol}(\Lambda_\mathbb{R})}.
\end{equation}
The value of $od(\mathcal{H})$ is always greater than or equal to one, and can only attain the unit if the columns of $\mathbf{H}$ are orthogonal to one another. We now use this metric to investigate how $N_R$ and $N_T$ influence the geometry of the mMIMO lattice and, therefore, how far from optimal ZF detection is.

\begin{figure}[t]
\centering
   \includegraphics[width=0.95 \columnwidth, clip=true, draft=false]{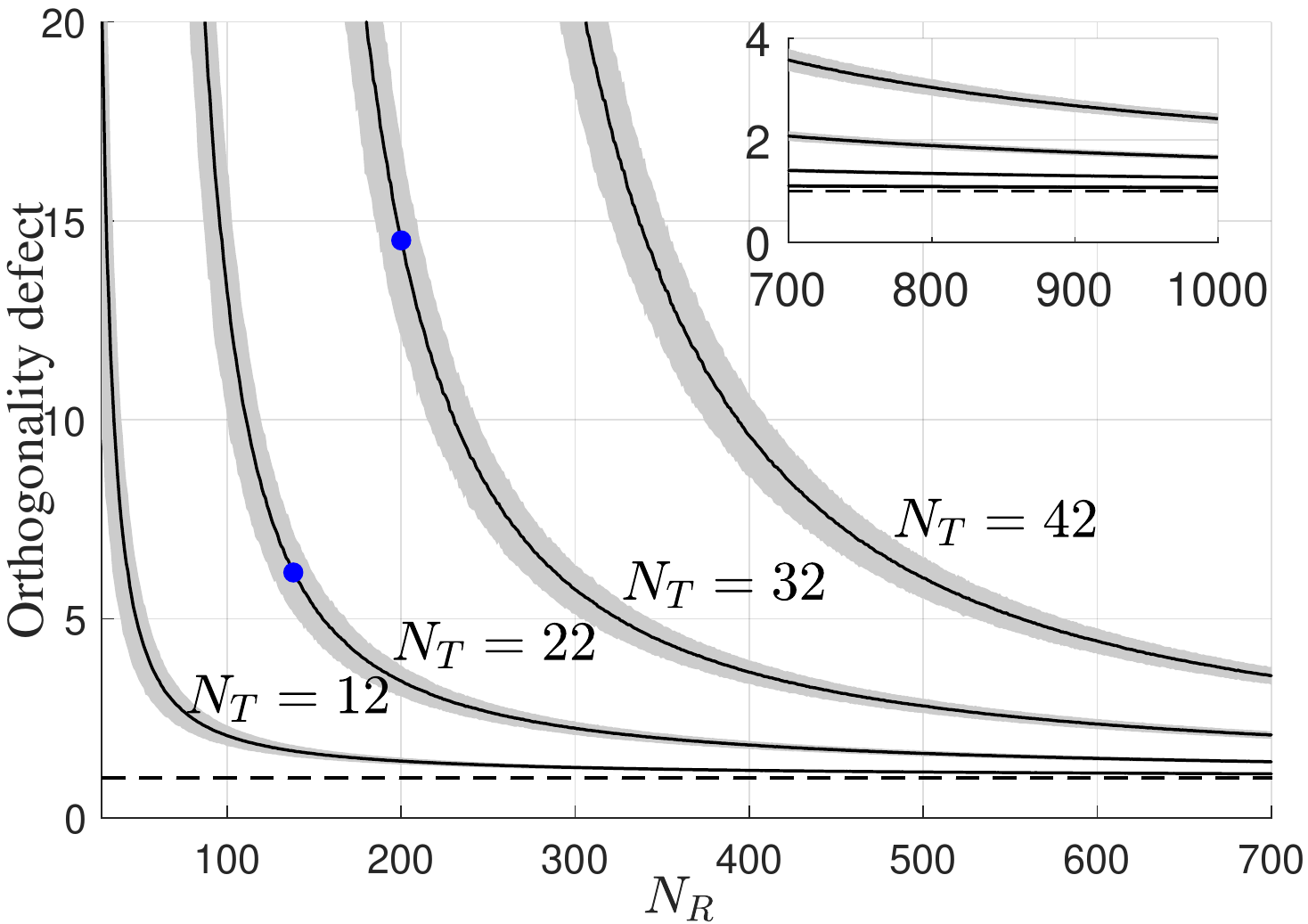}
\caption{Evolution of $od(\mathcal{H})$ as a function of the number of receive antennas. The blue dots indicate the operating points of the two systems that will be assessed in Section~\ref{sec:results} that are closer to (but still far) from the PCH regime. The shaded region corresponds to two standard deviations of $od$.}
\label{Orthogonality defect}
\end{figure}

Fig. \ref{Orthogonality defect} shows how the OD evolves with $N_R$, for different values of $N_T$.
The figure shows the domain of $N_R$ of more practical significance and an overlaid graph depicts the OD asymptotic convergence to the unit value when the number of receive antennas tends to infinity.
The OD is assessed by generating random samples of $\mathcal{H}$, with its real entries drawn from $\mathcal{N}(0, \frac{1}{2})$. This corresponds to generating $2N_T$ random Gaussian vectors in a vector space of $N_R$ (real) dimensions, with the dimension of the vector space being much larger than the number of random vectors drawn (i.e., $N_R>>N_T$). When this happens, those vectors are mutually orthogonal with high probability. As expected, larger $N_T$ necessitates having a larger $N_R$ in order to maintain the same $od(\mathcal{H})$ value, and as $N_R$ increases, the column vectors of $\mathcal{H}$ tend to be mutually orthogonal. 

\subsection{Perfect channel hardening lower-bound} 
\label{sec:CG}

There is one specific (and ideal) circumstance in which noise amplification is prevented: when all the column vectors in $\mathbf{H}$ are mutually orthogonal. This occurs when $N_T$ is fixed and $N_R \rightarrow \infty$, leading to the so-called \textit{channel hardening effect} \cite{bjornson2017massive}. For a geometric interpretation of this property, one could consider $N_T$ random Gaussian vectors living in a finite $N_R$-dimensional space. With high probability any pair of the $N_T$ vectors will be orthogonal to each other. With $N_R \rightarrow \infty$, this probability becomes 1. Let us consider a finite $N_R$ and the special case of a channel matrix where an \textit{ideal} MIMO channel is formed, i.e., a case where all columns of $\mathbf{H}$ are mutually orthogonal.
In this case, the Gram matrix, which comprises all inner products $\mathbf{h}^H_{i}\mathbf{h}_{j}$, $i=1,\dots\ N_R$, $j=1,\dots\ N_T$, becomes a diagonal matrix of the form:
\begin{equation}\label{eq_Gram}
    \mathbf{G}=\left[
    \begin{array}{ccccc}
    \|\mathbf{h}_{1}||^2     &    &    \text{\huge0} \\
     &       \ddots      &      \\
      \text{\huge0}  &  & \|\mathbf{h}_{N_T}||^2
    \end{array}
    \right]= N_R \mathbf{I}_{N_T},
\end{equation}
given that $\|\mathbf{h}_{j}||^2 =\sum_{i=1}^{N_R} |\mathbf{h}_{ij}|^2 = N_R$, for all the $N_T$ vectors.

Therefore, by replacing \eqref{eq_Gram} in \eqref{eq:R_u} one gets that the autocorrelation of the noise after ZF, in the case of PCH, is
\begin{equation}
    \mathbf{R_u}=  \frac{\sigma_\text{n}^2}{N_R} \mathbf{I}_{N_T}
\end{equation}
Finally, the power of $\mathbf{u}$ is
\begin{equation}
    \|\mathbf{u}\|^2=\Tr \left( \mathbf{R_u} \right) = \frac{\sigma_\text{n}^2 N_T}{N_R}.
\end{equation}

It is now possible to establish the equivalent channel model if the $N_T \times N_R$ mMIMO configurations were to attain PCH at those (finite) dimensions:
\begin{equation}\label{eq:CH}
\mathbf{H^+} \mathbf{y}_\text{s}= \sqrt{\frac{\mathrm{snr}}{N_T}} \mathbf{I}_{N_T} \mathbf{x}_\text{s}+\mathbf{u},
\end{equation}
In this scenario, one has $N_T$ independent parallel channels, where the effective noise becomes again a vector of independent Gaussian entries. Each of these $N_T$ components of $\mathbf{u}$ has power $|u_i|^2=\frac{\sigma_\text{n}^2}{N_R}$, shedding light on the benefit of having a larger receiver array: with $N_R \rightarrow \infty$ there is a regression to the mean, and the effective noise power vanishes.

Note that the $\mathrm{snr}$ in \eqref{eq:CH} is the \textit{input} SNR, at each receive antenna before any baseband processing takes place.
This asymptotic regime leads to a uniform \textit{post-processing} $\mathrm{snr}^{(ZF)}_i$ across the $N_T$ spatially multiplexed layers. In that limit, the reliability of all symbols is equal, and therefore sorting would bring no benefit.

%\begin{figure}[t]
%\centering
%\includegraphics[width=0.6\columnwidth]{Figures/soft_symbols.pdf}
%\caption{Hard detection of the received symbol.}
%\label{fig:detection}
%\end{figure}

\section{Symbol-level GRAND with antenna-sorting}
\label{sec:symbolLevelGRAND}

The application of symbol-level GRAND to a mMIMO system is not straightforward, given that it was originally proposed for a SISO block Rayleigh fading channel \cite{Chatzigeorgiou_Monteiro_2023}. Taking in consideration the analysis made in the previous section, this section outlines the principles of symbol-level GRAND and describes how it can be integrated into the mMIMO setup by incorporating soft information emanating from the ZF detector.

\subsection{Sorting error patterns guided by the constellation structure}

\begin{figure}[t]
    %\begin{center}
   \centering
    \includegraphics[width=0.65 \columnwidth, clip=true, draft=false]{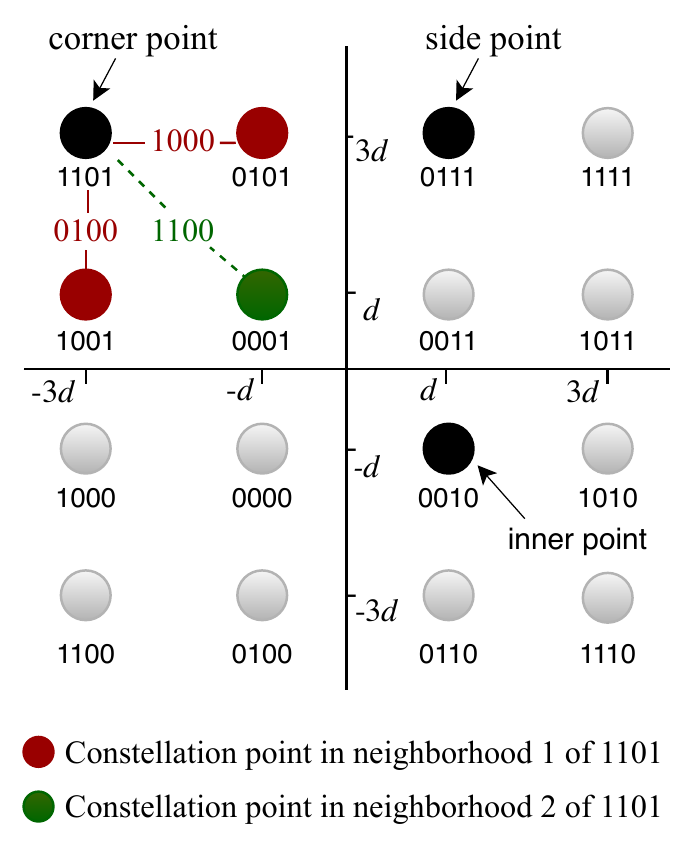}\textbf{}
   % \end{center}
   \caption{Examples of corner, side, and inner points in 
a 16-QAM constellation (all shown in black). The nearest neighbors of the symbol carrying string 1101 are shown in dark red (neighborhood 1), and the next-nearest neighbor of 1101 is shown in dark green (neighborhood 2). The error strings between 1101 and each neighbor are also depicted using the same color-coding scheme.}
    \label{fig:symbol-level}
\end{figure}

As previously stated, the $n$-bit codeword $\mathbf{x}_\text{b}$ can be written as a sequence of $L$ strings, i.e., $\mathbf{x}_\text{b}=\left[\mathcal{S}(s_i)\right]_{i=1}^L$, where $s_i$ is a symbol of the $M$-QAM constellation that corresponds to string $\mathcal{S}{(s_i)}$ of length $\log_2(M)$ bits. Fig.~\ref{fig:symbol-level} shows the bit string associated with each symbol of the $M$-QAM constellation when Gray mapping is used. The Euclidean distance between two adjacent symbols along one dimension is $2d$, where $d$ is given by \cite{lu1999m} \cite{cho2002general}:
\begin{equation}\label{eq:Euclidean_distance}
d=\sqrt{\frac{3}{2(M-1)}}.
\end{equation}
This Euclidean distance ensures that the average energy per $M$-QAM symbol is $\mathbb{E}\left\{|s_i|^2\right\}=1$. Observe that each symbol $s_i$ or, equivalently, bit string $\mathcal{S}{(s_i)}$, in the constellation is surrounded by symbols that belong to one of two neighborhoods: neighborhood~$1$, denoted by $\mathcal{N}_1(\mathcal{S}(s_i))$ and associated with symbols at a distance $2d$ from $s_i$, and neighborhood~$2$, denoted by $\mathcal{N}_2(\mathcal{S}(s_i))$ and associated with symbols at a distance $2\sqrt{2}d$ from $s_i$, as illustrated in Fig. \ref{fig:symbol-level}. Examples of symbols, referred to by the bit strings they carry, and their neighborhoods are provided in Table \ref{Neighbor}.

\begin{table}[t]
    \centering
\renewcommand{\arraystretch}{1.1}
 \caption{Neighborhoods of the symbols in the bottom-right quadrant of the $M$-QAM constellation shown in Fig.~\ref{fig:symbol-level}.}
%\begin{tabular}{|p{0.5cm}|p{3cm}|p{3cm}|}
\begin{tabular}{|p{0.5cm}|p{2.7cm}|p{2.7cm}|}
\hline 
\multicolumn{1}{|c}{$\mathcal{S}(s_i)$} & \multicolumn{1}{|c}{ $\mathcal{N}_1(\mathcal{S}(s_i))$} & \multicolumn{1}{|c|}{$\mathcal{N}_2(\mathcal{S}(s_i))$} \\
\hline 
\hline 1110 & 1010, 0110 & 0010 \\
\hline 1010 & 1011, 1110, 0010 & 0011, 0110 \\
\hline 0110 & 0010, 1110, 0100 & 0000, 1010 \\
\hline 0010 & 0000, 1010, 0011, 0110 & 0001, 0100, 1011, 1110 \\
\hline
\end{tabular}
    \label{Neighbor}
\end{table}

Modulo-2 addition of $\mathcal{S}({s}_i)$ with all elements of neighborhoods 1 and 2 generates the sets of \textit{error strings} $\mathcal{E}_1\left(\mathcal{S}(s_i)\right)$ and $\mathcal{E}_2\left(\mathcal{S}(s_i)\right)$, respectively. For instance, the bit strings $0101$ and $1001$ make up neighborhood 1 of $\mathcal{S}(s_i)=1101$ in Fig.~\ref{fig:symbol-level}; these two bit strings produce the error strings $1101\oplus0101 = 1000$
and $1101\oplus1001 = 0100$, therefore $\mathcal{E}_1(1101) =\{1000,0100\}$. Neighborhood 2 of 1101 consists only of bit string $0001$; hence $\mathcal{E}_2(1101) = \{1100\}$. Owing to Gray coding, the Hamming weight of all elements in $\mathcal{E}_1(s_i)$ is $1$, whereas the Hamming weight of all elements in $\mathcal{E}_2(s_i)$ is $2$, for any ${s}_i\in\mathcal{A}$. The position of a symbol $s_i$ in the constellation affects the cardinalities of $\mathcal{E}_1\left(\mathcal{S}(s_i)\right)$ and $\mathcal{E}_2\left(\mathcal{S}(s_i)\right)$. As seen in Fig.~\ref{fig:symbol-level} and can be inferred from Table~\ref{Neighbor}, if a symbol $s_i$ occupies a corner, side or inner point of the square $M$-QAM constellation, then $\mathcal{E}_1\left(\mathcal{S}(s_i)\right)$ has 2, 3 or 4 elements, and $\mathcal{E}_2\left(\mathcal{S}(s_i)\right)$ has 1, 2 or 4 elements, respectively.

At the receiver, the demodulator outputs $\mathbf{y}_\text{b}$, which can be expressed as a sequence of $L$ strings, that is, $\mathbf{y}_\text{b}=\left[\mathcal{S}(\tilde{s}_i)\right]_{i=1}^L$, as explained in Section~\ref{sec:ZFandGRAND}. Bit-level GRAND keeps generating and testing error patterns $\hat{\mathbf{e}}_\text{b}$ in descending order of likelihood until an error pattern that satisfies $\mathbf{y}_\text{b}\oplus \hat{\mathbf{e}}_\text{b} \in\mathcal{C}$ is identified. The likelihood of each error pattern is assumed to be a monotonically decreasing function of its Hamming weight.

\begin{table}[t]
\centering
\caption{Examples of error patterns with structure $[L_1\; L_2]$ for $M=16$ and $n=16$.}
\renewcommand{\arraystretch}{0.2}
%\begin{tabular}{|p{1cm}|p{6cm}|p{1.4cm}|}
\begin{tabular}{|p{1cm}|p{5.1cm}|p{1.3cm}|}
\hline
\makecell[c]{Structure \\ $[L_1\; L_2]$} & \vspace*{\fill} \makecell[c]{Examples of error patterns having structure \\ $[L_1\; L_2]$} & \makecell[c]{Weight\\($L_1+2L_2$)}\\
\hline \hline
\makecell[c]{$[2\; 0]$} & \makecell[c]{Example 1: $\textcolor{blue}{0010}-\textcolor{blue}{1000}-0000-0000$\\Example 2: $0000-\textcolor{blue}{0001}-\textcolor{blue}{0001}-0000$\\Example 3: $\textcolor{blue}{0100}-0000-0000-\textcolor{blue}{0010}$} & \makecell[c]{2} \\
\hline
\makecell[c]{$[0\; 1]$} & \makecell[c]{Example 1: $0000-\textcolor{red}{0011}-0000-0000$\\Example 2: $0000-0000-\textcolor{red}{1001}-0000$\\Example 3: $\textcolor{red}{1100}-0000-0000-0000$}& \makecell[c]{2} \\
\hline
\makecell[c]{$[1\; 1]$} & \makecell[c]{Example 1: $\textcolor{red}{0101}-\textcolor{blue}{0001}-0000-0000$\\Example 2: $\textcolor{blue}{0100}-0000-0000-\textcolor{red}{1100}$\\Example 3: $0000-0000-\textcolor{red}{0110}-\textcolor{blue}{1000}$}& \makecell[c]{3} \\
\hline
\makecell[c]{$[2\; 1]$} & \makecell[c]{Example 1: $0000-\textcolor{blue}{1000}-\textcolor{red}{0011}-\textcolor{blue}{0100}$\\Example 2: $\textcolor{red}{0110}-\textcolor{blue}{0001}-\textcolor{blue}{0001}-0000$\\Example 3: $\textcolor{blue}{0010}-0000-\textcolor{blue}{0100}-\textcolor{red}{1010}$}& \makecell[c]{4} \\
\hline
\end{tabular}
\label{error pattern}
\end{table}

In symbol-level GRAND, the  prerequisite for $\mathbf{y}_\text{b}\oplus \hat{\mathbf{e}}_\text{b} \in\mathcal{C}$ remains in place but is stated as $\left[\mathcal{S}(\tilde{s}_i)\oplus \hat e_i\right]_{i=1}^L \in \mathcal{C}$, where $\hat e_i$ is the $i$-th error string of the error pattern $\hat{\mathbf{e}}_\text{b}=\left[\hat{e_i}\right]_{i=1}^L$. Given the structure of the $M$-QAM constellation, $\hat e_i$ will, with high probability, belong to either $\mathcal{E}_1(\mathcal{S}(\tilde{s}_i))$ or $\mathcal{E}_2(\mathcal{S}(\tilde{s}_i))$, or it will be a string of $\log_2(M)$ zeros, denoted by $\mathbf{0}$. Differently from bit-level GRAND, symbol-level GRAND does not generate and verify each realization of $\hat{\mathbf{e}}_\text{b}$ for increasing Hamming weight. It creates and queries realizations of $\hat{\mathbf{e}}_\text{b}$ that are composed of error strings that are more likely to occur, i.e., if $\mathcal{S}(\tilde{s}_i)$ is the $i$-th detected and potentially erroneous bit string, then $\hat e_i \in \mathcal{E}_1\left(\mathcal{S}(\tilde{s}_i)\right) \cup \mathcal{E}_2\left(\mathcal{S}(\tilde{s}_i)\right)\cup \{\mathbf{0}\}$. Hereafter, for simplicity, we say that an error string $\hat e_i$ is of \textit{type} $\mathcal{E}_j$ if $\hat e_i\in\mathcal{E}_j\left(\mathcal{S}(\tilde{s}_i)\right) $ for $j = 1, 2$.

Table \ref{error pattern} contains examples of error patterns for $M=16$ and $n=16$. Each error pattern consists of \mbox{$L=n/\log_2(M)=4$} error strings, each having length $\log_2(M)=4$ bits. For clarity, error strings in an error pattern have been separated by dashes. Following the notation of \cite{Chatzigeorgiou_Monteiro_2023}, the structure of error patterns has been denoted by $[L_1\; L_2]$, where $L_1$ is the number of type-$\mathcal{E}_1$ error strings (displayed in blue), $L_2$ is the number of type-$\mathcal{E}_2$ error strings (displayed in red), and $L-L_1-L_2$ is the number of error strings that contain only zeros (displayed in black). Recall that type-$\mathcal{E}_1$ error strings have weight $1$, whereas type-$\mathcal{E}_2$ error strings have weight $2$. Thus, the weight of an error pattern that consists of $L_1$ error strings of type $\mathcal{E}_1$ and $L_2$ error strings of type $\mathcal{E}_2$ is $L_1+2L_2$.

The probability that an error pattern with structure $[L_1 \ L_2]$ has occurred was derived in~\cite{Chatzigeorgiou_Monteiro_2023} and is used by symbol-level GRAND to arrange and test error patterns in descending order of likelihood. In \cite{Chatzigeorgiou_Monteiro_2023}, block fading was considered, thus all received symbols were affected by the same fading coefficient and, consequently, had the same reliability. For this reason, symbol-level GRAND assumes that error patterns with the same structure $[L_1 \ L_2]$ have the same probability of occurrence. However, received symbols have different reliabilities in the considered mMIMO setup. As explained in the following section, CSI can be used to guide symbol-level GRAND on how to prioritize error patterns of the same structure.

\subsection{Sorting error patterns guided by CSI} \label{subsec:sorting}

In a non-ideal mMIMO scenario with $od(\mathcal{H})>1$, the reliability of spatial streams may greatly differ among them. At the receiver, a first processing block should implement the antenna sorting discussed in Section \ref{sec:sorting}.

\begin{figure}[t]
   \centering
   \subfigure[]
    {\includegraphics[width=0.9\columnwidth, clip=true, draft=false]{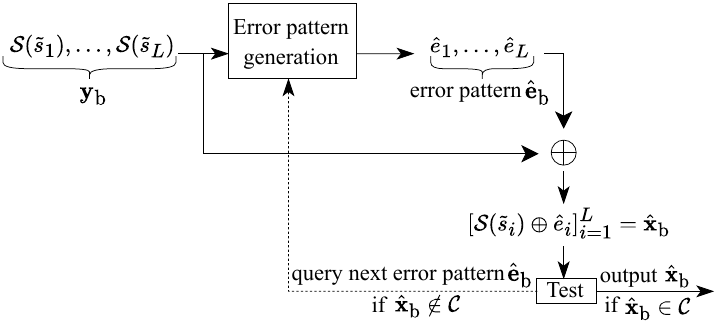}}\quad
   \subfigure[]
   {\includegraphics[width=0.9\columnwidth, clip=true, draft=false]{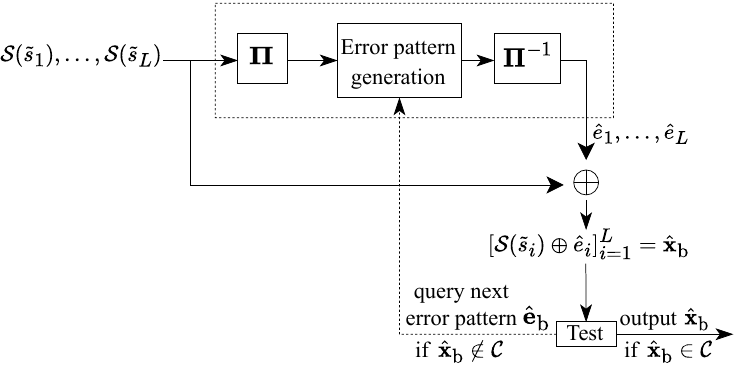}}
 \caption{Types of symbol-level GRAND: (a) symbol-level GRAND proposed in \cite{Chatzigeorgiou_Monteiro_2023}, and (b) symbol-level GRAND with antenna sorting.}
   \label{permutations}
 \end{figure}

Without loss of generality, we will discuss the case with $N_c=1$. This can be accomplished by inserting a permutation matrix $\mathbf{\Pi}$, which is a binary matrix whose columns are all columns of the identity $\mathbf{I}$ but placed in a different order. As it is well known, a permutation matrix is always an orthogonal matrix, and its inverse is its transpose: $\mathbf{\Pi}^{-1}= \mathbf{\Pi}^T$. These two matrices can be added before and after symbol-level GRAND, as presented in Fig. \ref{permutations}.
The $g_i$ gains in (\ref{SNR-gain}) are sorted in ascending order and the corresponding permutation matrix $\mathbf{\Pi}$ is created. The permuted symbols $\tilde{\mathbf{x}}_\text{s}^{(\Pi)}=\tilde{\mathbf{x}}_\text{s}\mathbf{\Pi}$ are fed to symbol-level GRAND, which will test error patterns in decreasing order of likelihood, which is additionally helped by this sorting mechanism.

%  \begin{figure*}[t]
%   \centering
%    \subfigure[]
%     {\includegraphics[width=0.48\columnwidth, clip=true, draft=false]{Figures/Without_antenna_sorting.pdf}}\quad
%    \subfigure[]
%      {\includegraphics[width=0.48\columnwidth, clip=true, draft=false]{Figures/With_antenna_sorting.pdf}}
%   \caption{Types of symbol-level GRAND: (a) symbol-level GRAND proposed in \cite{Chatzigeorgiou_Monteiro_2023}, and (b) symbol-level GRAND with antenna sorting.}
%   \label{permutations}
% \end{figure*}  

\begin{figure}[t]
    %\begin{center}
    \centering
    \includegraphics[width=0.57 \columnwidth, clip=true, draft=false]{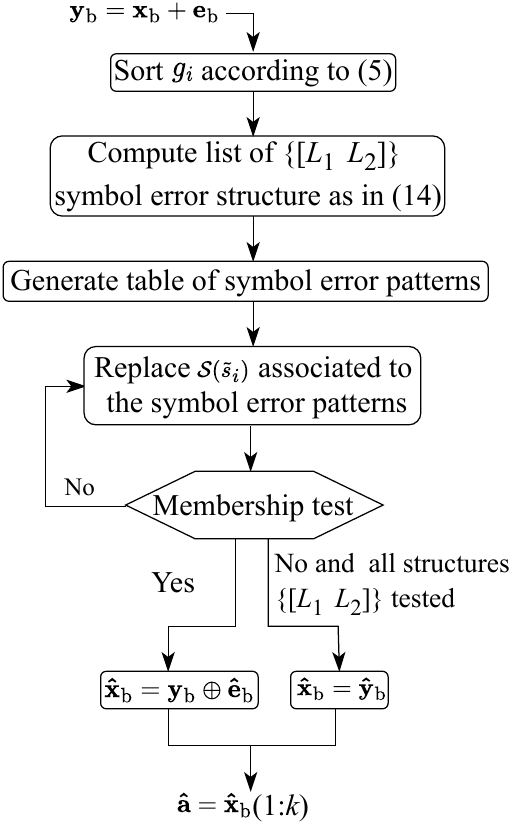}

   % \end{center}
   \caption{Symbol-level GRAND with antenna sorting after the demodulation and detection stages. (The final output of the algorithm makes use of MATLAB notation.)}
    \label{fig:SGRAND}
\end{figure}

In the general case with $N_c>1$ codewords per MIMO transmission, the set of the $g_i$, for $i=1, \cdots, N_T=N_cL$, is partitioned in $N_c$ subsets and an independent sorting process is applied to each one of those subsets. Note that, for a faster overall decoding time, these sorting processes can be implemented in parallel. Afterwards, each subset of $L$ sorted symbols is passed on to the symbol-level GRAND, which independently decode each one of these $N_c$ codewords. Likewise the sorting procedures, the decoding of each codeword can be performed in parallel, if further reduction of decoding latency is paramount. This may be done at the cost of having multiple symbol-level GRAND processors.

The flowchart of the proposed symbol-level GRAND with antenna sorting is presented in Fig.~\ref{fig:SGRAND} (for the $N_c=1$ case). The probability expression for error patterns with structure $[L_1 \ L_2]$ is discussed in the next section in the context of a mMIMO system operating in the PCH limit. 

%\begin{figure}[ht]
%    \centering
%    \includegraphics[width=1 \columnwidth, clip=true, draft=false]{Figures/16-QAM_extended_compact.pdf}
%    \caption{Example of the extended region considered in calculating the probability that 0110 was transmitted given that 0010 is received.}
%    \label{fig:16-QAM_extended_compact}
%\end{figure}

\section{Analysis in the perfect channel hardening limit}
\label{sec:analysis}

A closed-form approximation for the probability of an error pattern with structure $[L_1 \ L_2]$ was derived in \cite{Chatzigeorgiou_Monteiro_2023} for SISO block Rayleigh fading channels impaired by AWGN. The approximated probability of occurrence of an error pattern that consists of $L$ error strings, with $L_1$ of them being of type $\mathcal{E}_1$ and $L_2$ error strings being of type $\mathcal{E}_2$, is given by \cite{Chatzigeorgiou_Monteiro_2023}:
\begin{IEEEeqnarray}{l}
\label{expression}
P(L_1,L_2)\approx%
\nonumber%
\\
\sum_{L_\mathrm{c}+L_\mathrm{s}+L_\mathrm{i}=L}\!\binom{L}{L_\mathrm{c},L_\mathrm{s},L_\mathrm{i}}4^{L_\mathrm{c}+L_\mathrm{s}}\Bigl(\sqrt{M}-2\Bigr)^{L_\mathrm{s}+2L_\mathrm{i}}
\nonumber%
\\
\times\!\!\!\!\sum_{\substack{L_\mathrm{c,e}+L_\mathrm{s,e}+L_\mathrm{i,e}=\,L_1+L_2 \\ L_\mathrm{c,e}\leq L_\mathrm{c} \\ L_\mathrm{s,e}\leq L_\mathrm{s} \\ L_\mathrm{i,e}\leq L_\mathrm{i}}}\;\,\prod_{\substack{\forall\ell\in\mathcal{L} \\ \mathcal{L}=\{\mathrm{c,s,i}\}}}\!\!\!\binom{L_\ell}{L_{\ell,\mathrm{e}}}p^{L_\ell-L_{\ell,\mathrm{e}}}_{\ell,0}%
\nonumber%
\\
\times\!\!\!\!\sum_{\substack{L_\mathrm{c,e_1}+L_\mathrm{s,e_1}+L_\mathrm{i,e_1}=\,L_1 \\ L_\mathrm{c,e_1}\leq L_\mathrm{c,e} \\ L_\mathrm{s,e_1}\leq L_\mathrm{s,e} \\ L_\mathrm{i,e_1}\leq L_\mathrm{i,e}}}\;\prod_{\forall\ell\in\mathcal{L}}\!\binom{L_{\ell,\mathrm{e}}}{L_{\ell,\mathrm{e_1}}}p^{L_{\ell,\mathrm{e_1}}}_{\ell,\mathrm{e_1}} p^{L_{\ell,\mathrm{e}}-L_{\ell,\mathrm{e_1}}}_{\ell,\mathrm{e_2}}.%
\end{IEEEeqnarray}
The set $\mathcal{L}=\{\mathrm{c}, \mathrm{s}, \mathrm{i}\}$ contains the indices that signify the three types of constellation points that symbols can occupy, i.e., corner ($\mathrm{c}$), side ($\mathrm{s}$), or inner ($\mathrm{i}$) points, as illustrated in Fig.~\ref{fig:symbol-level}. For a given constellation point $\ell \in \mathcal{L}$, an error string will either contain zeros, be of type $\mathcal{E}_1$, or be of type $\mathcal{E}_2$. The probabilities associated with these errors are, respectively, $p_{\ell, 0}$, $p_{\ell, \mathrm{e}_1}$ and $p_{\ell, \mathrm{e}_2}$, and expressions for them were presented in \cite{Chatzigeorgiou_Monteiro_2023} but are also listed in Table \ref{table_probabilities} for the sake of completeness. They are all functions of the halfway Euclidean distance $d^\prime$ between any two adjacent points along one dimension of the constellation diagram at the receiver. More details about the derivation of the probabilities in Table~\ref{table_probabilities} can be found in the Appendix. A relationship between $d^\prime$ and the Euclidean distance $d$, which is observed in the constellation diagram at the transmitter, e.g., see Fig.~\ref{fig:symbol-level}, can be obtained for PCH.

As explained in Section~\ref{sec:CG}, PCH is achieved when $N_T$ is fixed and $N_R\rightarrow\infty$, which essentially reduces the mMIMO channel into an equivalent non-fading AWGN channel. In this case, the ergodic SNR in \eqref{eq:ergodic_SNR} and the noise variance $\sigma^2_\text{n}=1$, which implies that the noise variance per dimension is $0.5$, can be used to obtain $d^\prime$ as follows:
\begin{equation}
 d^ {\prime}= d \, \frac{\sqrt{\mathrm{snr}}}{\sqrt{0.5}} = d\,\sqrt{2\,\mathrm{snr}} = \sqrt{\frac{3\,\mathrm{snr}}{M-1}},
\end{equation}
where $d$ is defined in \eqref{eq:Euclidean_distance}.

\begin{figure}[t]
    \centering
\includegraphics[width=1 \columnwidth, clip=true, draft=false]{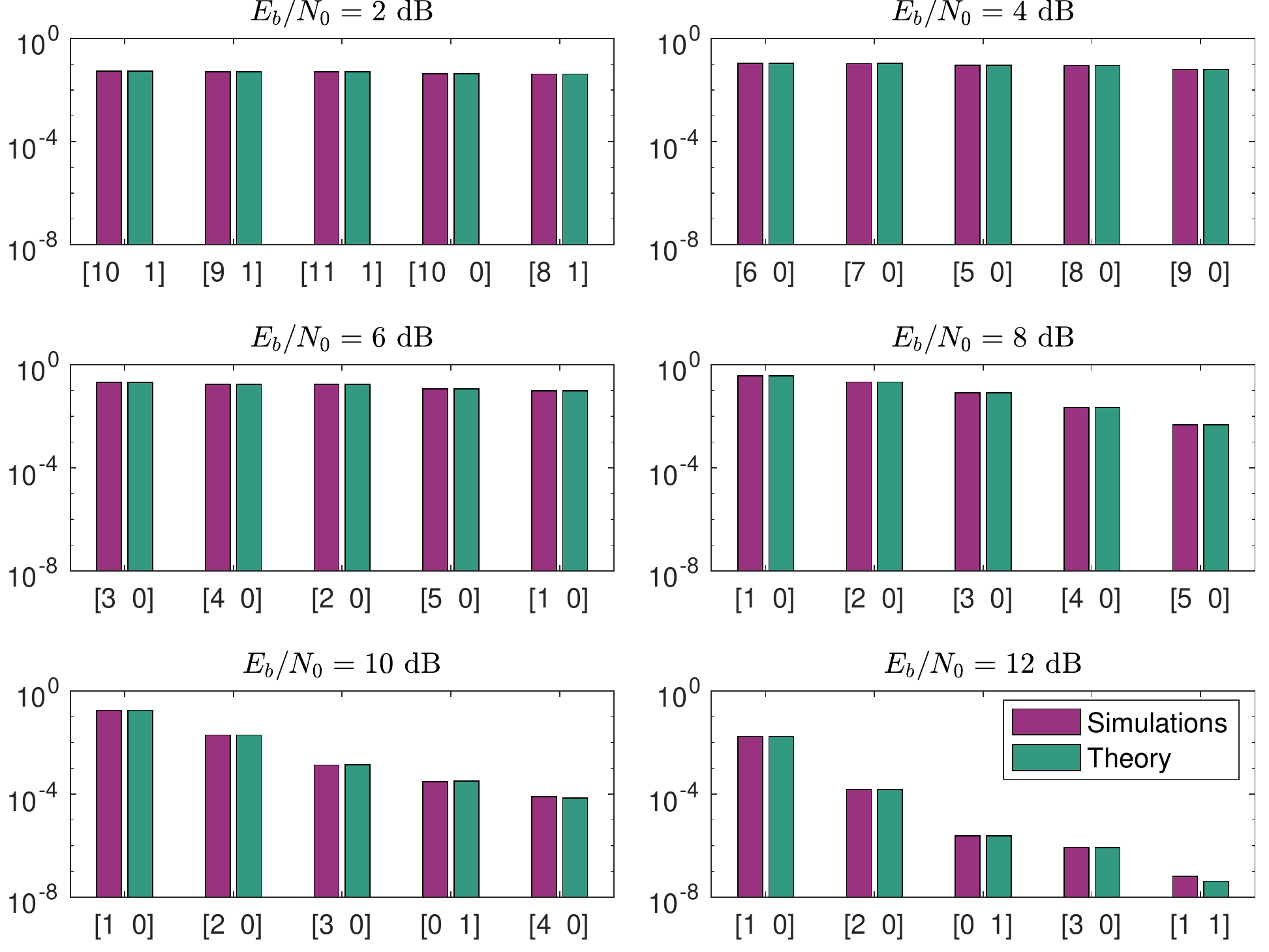}

    \caption{Structures of error patterns arranged in order of likelihood for $N_R\rightarrow\infty$ and different values of $E_b/N_0$ when $L$=32 symbols. A structure $[L_1 \ L_2]$ on the horizontal axis of any subplot represents error patterns containing $L_1$ type-$\mathcal{E}_1$ and $L_2$ type-$\mathcal{E}_2$ error strings, which occur with probability $P(L_1, L_2)$, shown on the vertical axis.}
    \label{fig:error_types}
\end{figure}

Fig. \ref{fig:error_types} shows the five most likely structures of error patterns for $N_R \rightarrow \infty$ and different values of $E_b/N_0$. Predicted probability values of $P(L_1, L_2)$, calculated from (\ref{expression}), are compared with measurements, obtained through simulations, for each $E_b/N_0$ value. The theoretical results match the simulation results, and confirm our hypothesis that the structure of an error pattern plays a more important role in the likelihood of that error pattern that its Hamming weight. As one would expect, error patterns that consist of only type-$\mathcal{E}_1$ error strings, i.e., with structure $[L_1 \ 0]$, become dominant at high $E_b/N_0$ values. Nevertheless, error patterns containing type-$\mathcal{E}_2$ error strings continue to appear among the most likely structures.

\begin{table}[t]
    \centering
        \caption{Expressions for the probability terms in (\ref{expression}). Function $Q(z) \triangleq(1 / \sqrt{2 \pi}) \int_z^{\infty} \exp \left(-t^2 / 2\right) dt$ is the tail distribution of the standard normal distribution. Variable $d^\prime$ is given by $d^{\prime}=\sqrt{3 \,\mathrm{snr} /(M-1)}.$}
    $$
\begin{array}{|l|}
\hline p_{\mathrm{c}, 0}=(1 / M)\left(1-Q\left(d^{\prime}\right)\right)^2 \\
p_{\mathrm{s}, 0}=(1 / M)\left(1-Q\left(d^{\prime}\right)\right)\left(1-2 Q\left(d^{\prime}\right)\right) \\
p_{\mathrm{i}, 0}=(1 / M)\left(1-2 Q\left(d^{\prime}\right)\right)^2 \\
\hline p_{\mathrm{c}, \mathrm{e}_1}=2(1 / M)\left(1-Q\left(d^{\prime}\right)\right) Q\left(d^{\prime}\right) \\
p_{\mathrm{s}, \mathrm{e}_1} \approx(1 / M)\left[2\left(1-Q\left(d^{\prime}\right)\right) Q\left(d^{\prime}\right)+\left(1-2 Q\left(d^{\prime}\right)\right) Q\left(d^{\prime}\right)\right] \\
p_{\mathrm{i}, \mathrm{e}_1} \approx 4(1 / M)\left(1-2 Q\left(d^{\prime}\right)\right) Q\left(d^{\prime}\right) \\
\hline p_{\mathrm{c}, \mathrm{e}_2}=(1 / M) Q^2\left(d^{\prime}\right) \\
p_{\mathrm{s}, \mathrm{e}_2} \approx 2(1 / M) Q^2\left(d^{\prime}\right) \\
p_{\mathrm{i}, \mathrm{e}_2} \approx 4(1 / M) Q^2\left(d^{\prime}\right) \\
\hline
\end{array}
$$    
    \label{table_probabilities}
\end{table}

\begin{figure}[t]
\centering
    %width=.7   width=1
        % \includegraphics[width=.7 \columnwidth, clip=true, draft=false]{Figures/LUT_snr.pdf}
    \includegraphics[width=1 \columnwidth, clip=true, draft=false]{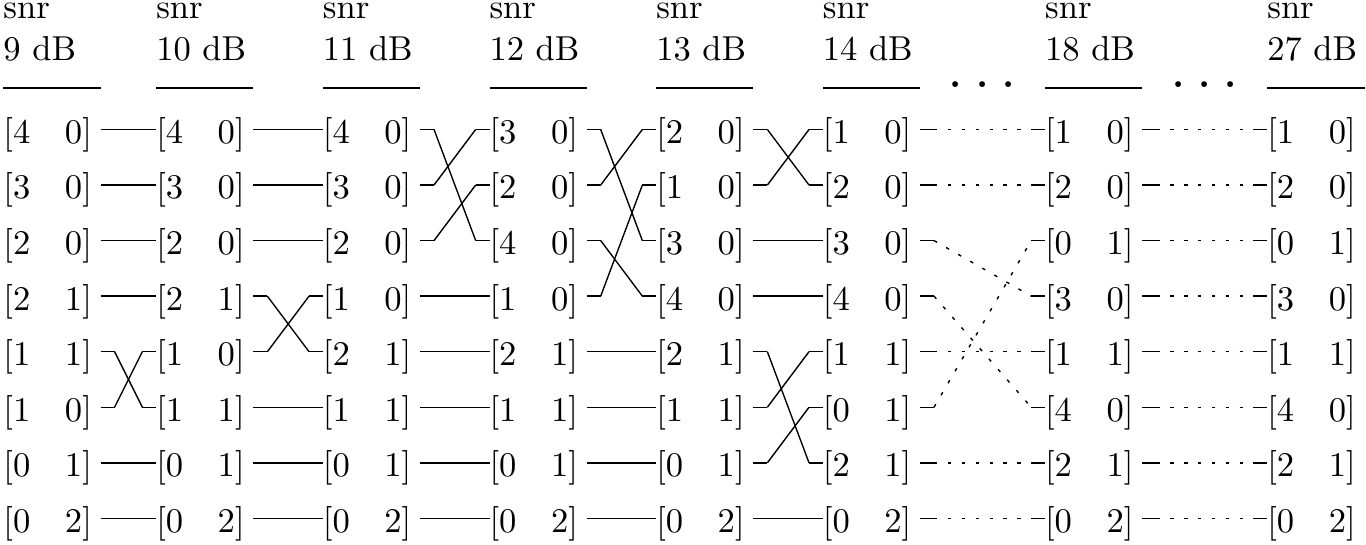}
\caption{Evolution of the ranking of the error structures, based on \eqref{expression}, for $N_R\rightarrow\infty$, $w_\mathrm{th}=4$ and an increasing value of $\mathrm{snr}$.}
\label{fig:LUT_per_SNR}
\end{figure}

As proven in \cite{Chatzigeorgiou_Monteiro_2023}, the worst-case number of error patterns tested by bit-level GRAND and symbol-level GRAND, until a codeword is estimated, is $M^L$ and $9^L$, respectively, although fewer tests are required on average \cite{Duffy2019}. This reduction in the search space -- and thus the complexity -- of symbol-level GRAND is achieved at the expense of a marginal increase in memory requirements. Antenna sorting, which is combined with symbol-level GRAND, has quasi-linear complexity when using the quick sort or the heap sort algorithms.

In order to reduce complexity, bit-level GRAND introduced the notion of the \textit{abandonment threshold}~\cite{Duffy2019}, denoted by $w_\text{th}$. 
Error patterns of increasing Hamming weight are tested but tests are abandoned and an error is declared, if  all error patterns of weight equal to or less than $w_\text{th}$ have been queried and a valid codeword has not been found. An abandonment threshold can also be used by symbol-level GRAND, so that error patterns of structure $[L_1 \ L_2]$ are queried only if their weight, given by $L_1+2L_1$, satisfies $0<L_1+2 L_2 \leq w_{\text{th}}$. This restriction further reduces the complexity of symbol-level GRAND and also decreases its memory requirements.

Fig.~\ref{fig:LUT_per_SNR} provides an example that clarifies the memory required by symbol-level GRAND. Lookup tables are presented side by side for $\mathrm{snr}$ values ranging from $9$~dB to $27$~dB in steps of $1$~dB. Each lookup table contains all possible structures for $w_\text{th}=4$, arranged in descending order of likelihood as determined by \eqref{expression}. One lookup table is required for each $\mathrm{snr}$ value in the range, since the ordering of the error structures depends on $\mathrm{snr}$. In this example, the ordering of the error structures does not change for $\mathrm{snr}$ values greater than $18$~dB, therefore lookup tables beyond $18$~dB can be omitted. In general, as described in \cite{Chatzigeorgiou_Monteiro_2023}, the memory required to store the lookup tables is $\lambda v \tau$ bits, where $\lambda$ is the number of bits needed to represent a structure $[L_1 \ L_2]$:
\begin{equation}
\label{eq:lambda_bits}
\lambda=\left\lceil\log _2\left(w_{\text {th}}+1\right)\right\rceil+\left\lceil\log _2\left(\left\lfloor w_{\text {th }} / 2\right\rfloor+1\right)\right\rceil\,\mathrm{bits},
\end{equation}
$v$ is the number of structures stored for each $\mathrm{snr}$ value, and $\tau$ is the number of $\mathrm{snr}$ values considered. Note that $\lfloor\cdot\rfloor$ and $\lceil\cdot\rceil$ in \eqref{eq:lambda_bits} denote the floor and ceiling functions, respectively. In the example presented in Fig.~\ref{fig:LUT_per_SNR}, a structure $[L_1 \ L_2]$ occupies $\lambda=5$ bits, each lookup table contains $v=8$ structures, and a total of $\tau=10$ lookup tables are needed to cover the range between $9$~dB and $18$~dB in steps of $1$~dB. Therefore, symbol-level GRAND will reserve $\lambda v \tau=400$ bits of memory space.

\section{Results}
\label{sec:results}

\begin{figure*}
    %\begin{center}
   \centering
\includegraphics[width=0.80 \textwidth]{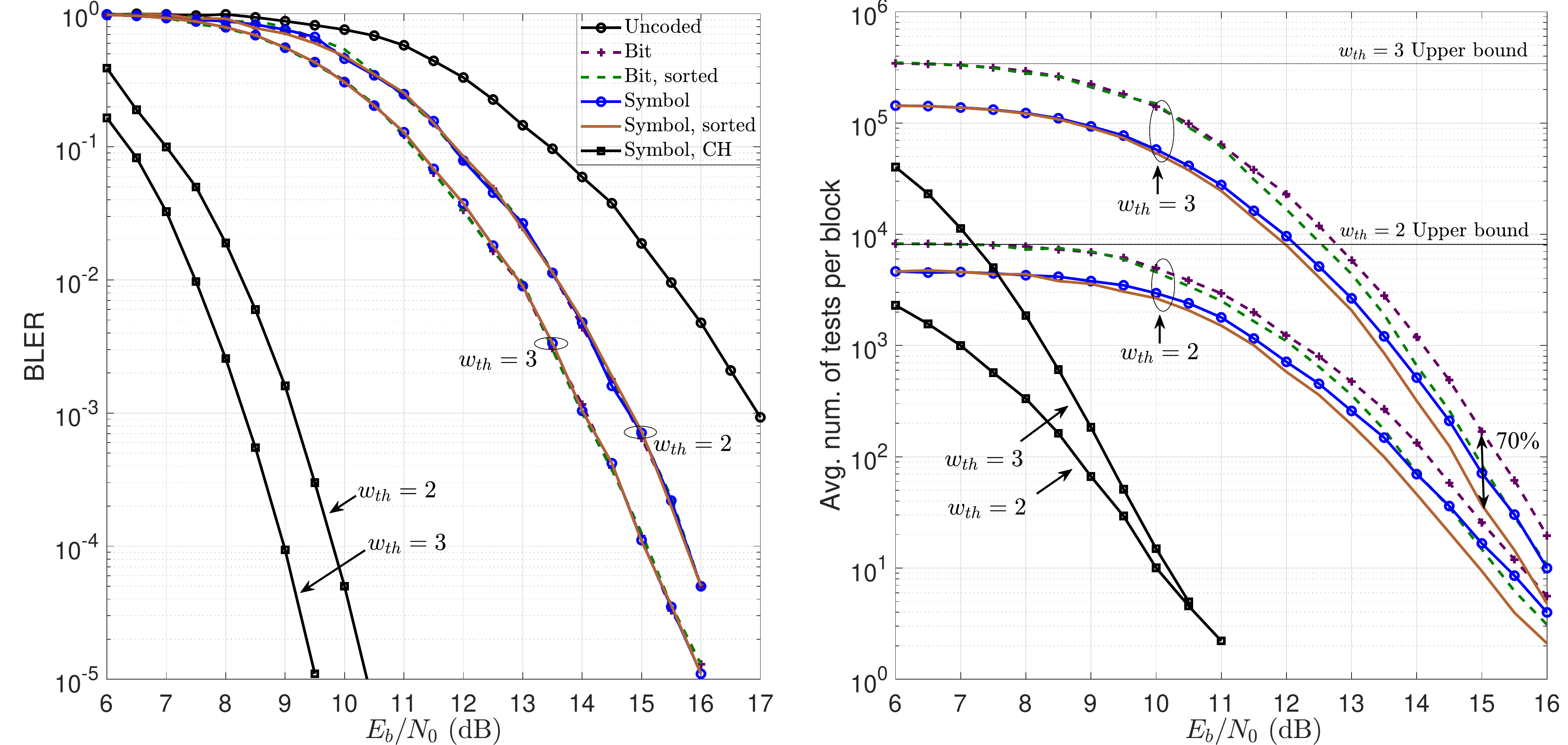}
%\includegraphics[width=1 \textwidth]{Figures/16QAM-N_R_50.pdf}
   % \end{center}
   \caption{BLER performance (left) and decoding complexity (right) for different thresholds $w_{th}=2,3$ in GRAND, using RLC (128,103), with $N_T=32$ and $N_R=50$, and 16-QAM. The corresponding PCH lower bounds are also plotted.}
   \label{fig:1}
\end{figure*}

\begin{figure*}
    %\begin{center}
    \centering
\includegraphics[width=0.80 \textwidth]{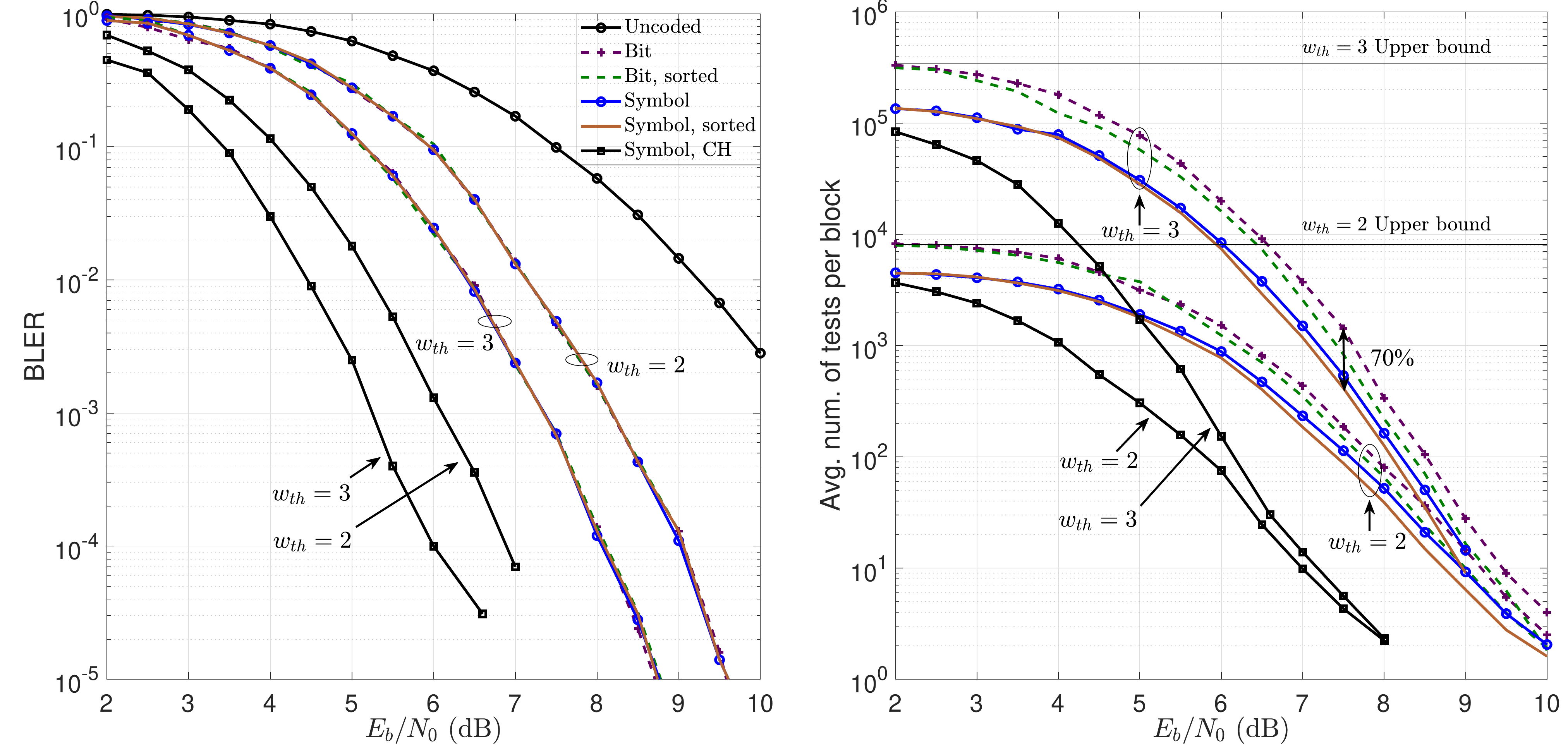}
%\includegraphics[width=1 \textwidth]{Figures/16QAM-N_R_100.pdf}
   % \end{center}
   \caption{BLER performance (left) and decoding complexity (right) for different thresholds $w_{th}=2,3$ in GRAND, using RLC (128,103), with $N_T=32$ and $N_R=100$, and 16-QAM. The corresponding PCH lower bounds are also plotted.}
   \label{fig:2}
\end{figure*}

\begin{figure*}
    %\begin{center}
    \centering
\includegraphics[width=0.80 \textwidth]{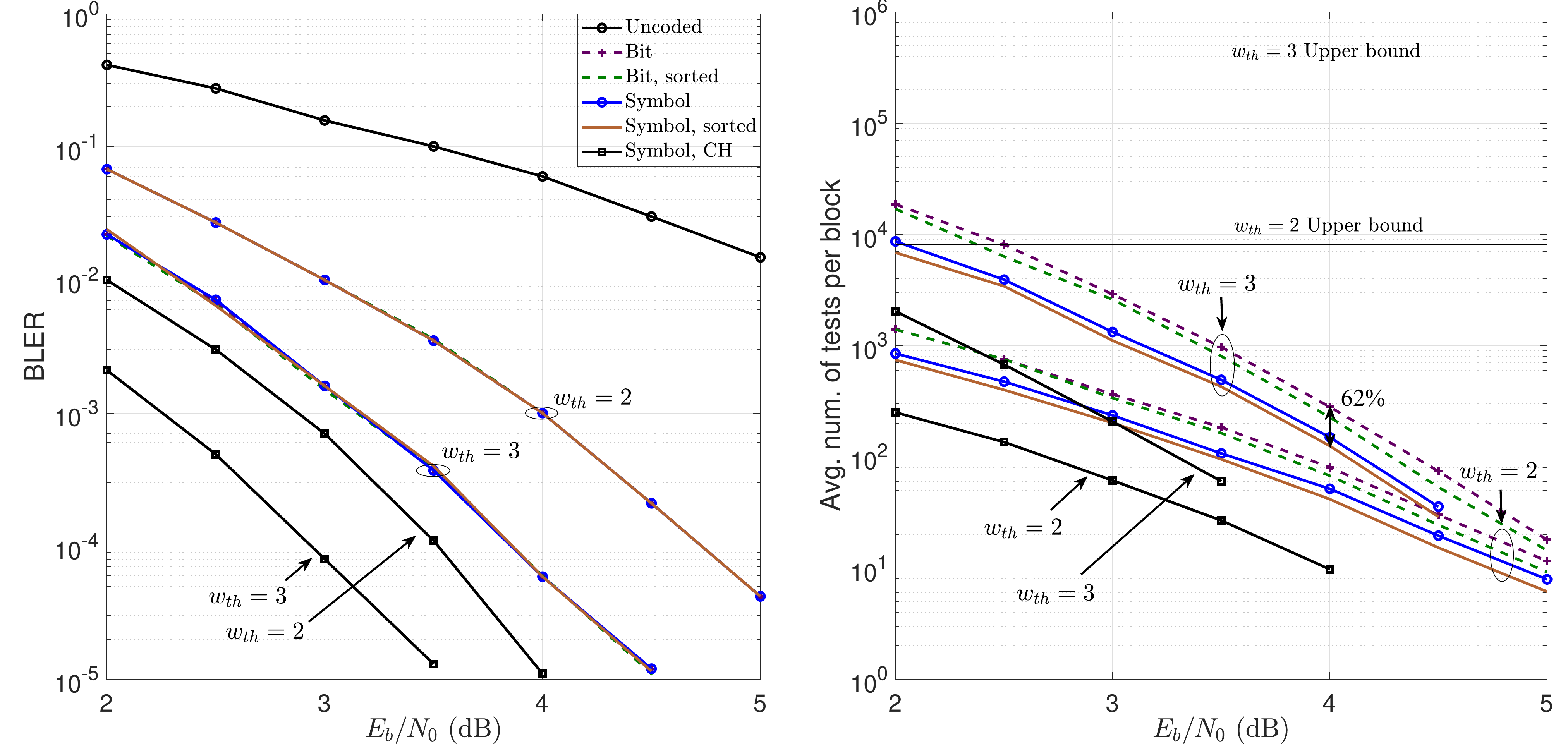}
%\includegraphics[width=1 \textwidth]{Figures/16QAM-N_R_200.pdf}
   % \end{center}
   \caption{BLER performance (left) and decoding complexity (right) for different thresholds $w_{th}=2,3$ in GRAND, using RLC (128,103), with $N_T=32$ and $N_R=200$, and 16-QAM. The corresponding PCH lower bounds are also plotted.}
   \label{fig:3}
\end{figure*}

\begin{figure*}
    %\begin{center}
   \centering
\includegraphics[width=0.8 \textwidth]{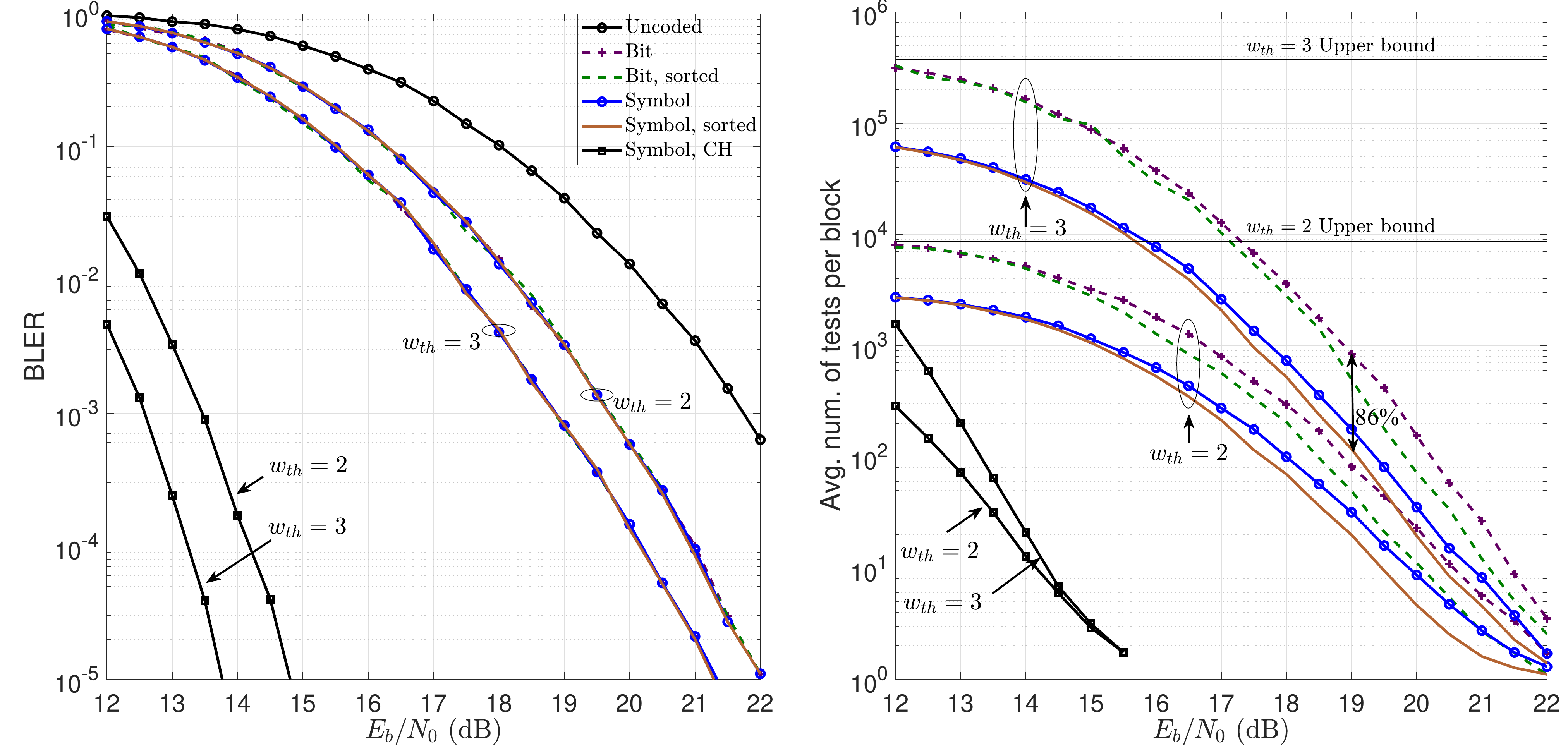}
%\includegraphics[width=1 \textwidth]{Figures/64QAM-N_R_34.pdf}
   % \end{center}
   \caption{BLER performance (left) and decoding complexity (right) for different thresholds $w_{th}=2,3$ in GRAND, using RLC (132,106), with $N_T=22$ and $N_R=34$, and 64-QAM. The corresponding PCH lower bounds are also plotted.}
   \label{fig:4}
\end{figure*}

\begin{figure*}
    %\begin{center}
    \centering
\includegraphics[width=0.8 \textwidth]{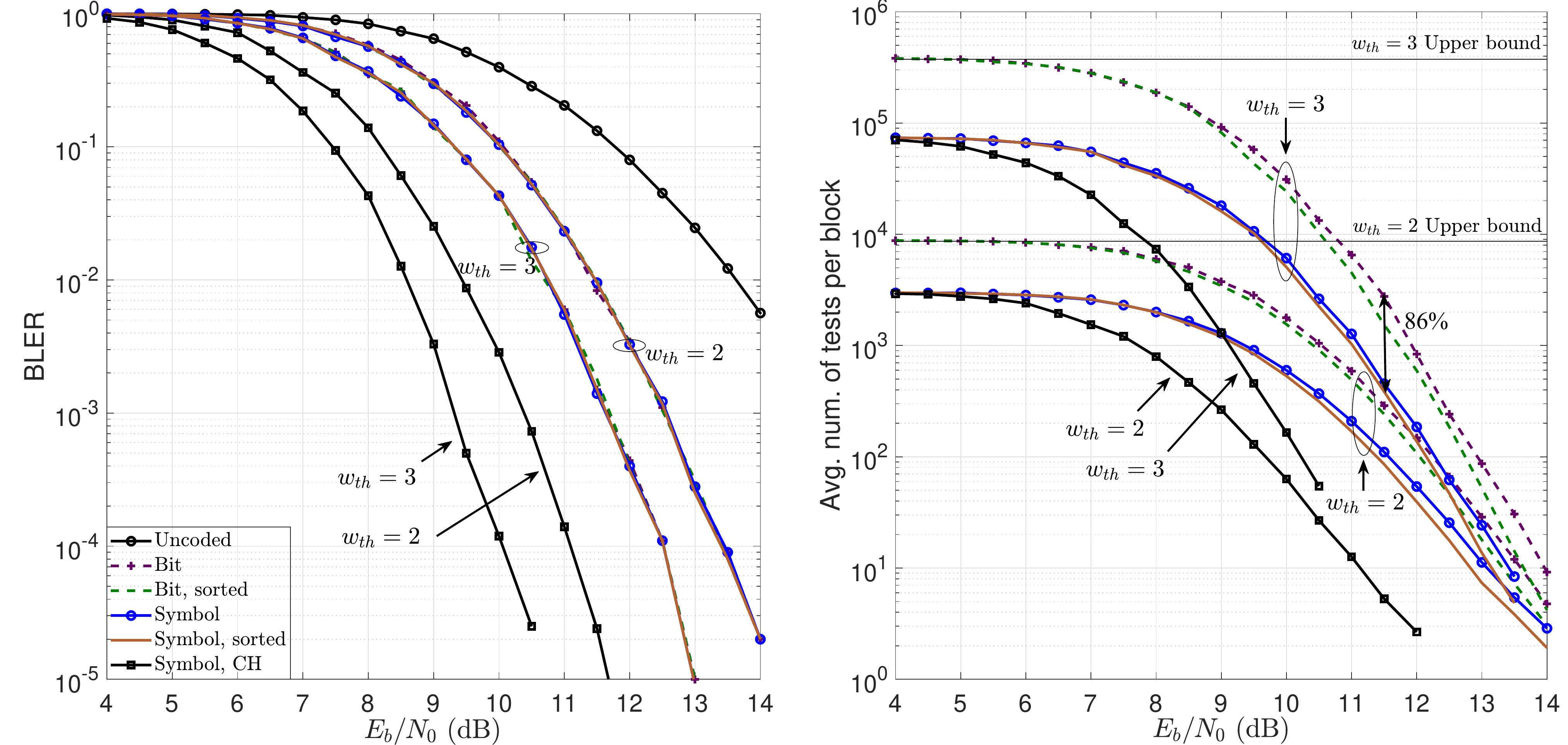}
%\includegraphics[width=1 \textwidth]{Figures/64QAM-N_R_69.pdf}
   % \end{center}
   \caption{BLER performance (left) and decoding complexity (right) for different thresholds $w_{th}=2,3$ in GRAND, using RLC (132,106), with $N_T=22$ and $N_R=69$, and 64-QAM. The corresponding PCH lower bounds are also plotted.}
   \label{fig:5}
\end{figure*}

\begin{figure*}
    %\begin{center}
    \centering
\includegraphics[width=0.8 \textwidth]{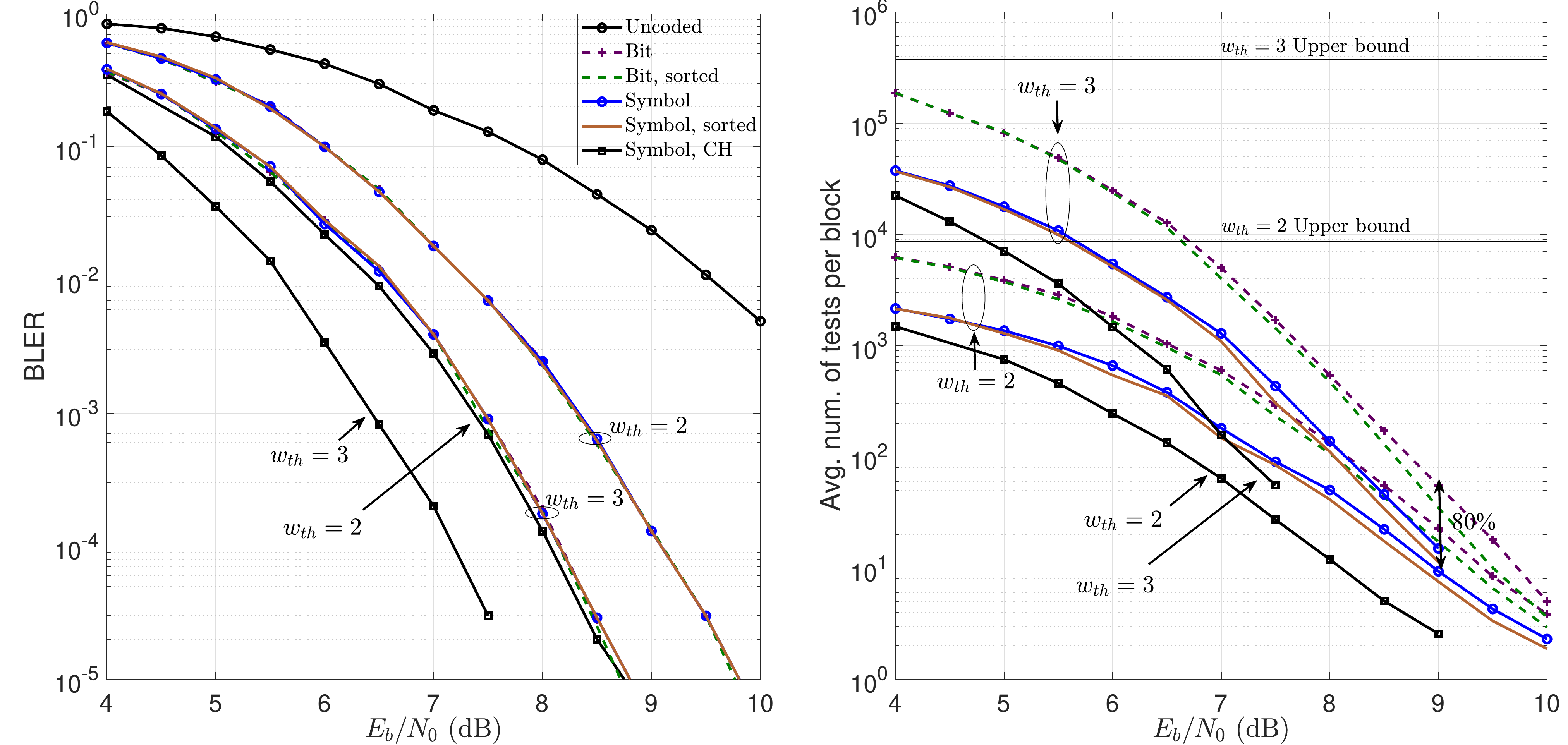}
%\includegraphics[width=1 \textwidth]{Figures/64QAM-N_R_138.pdf}
   % \end{center}
   \caption{BLER performance (left) and decoding complexity (right) for different thresholds $w_{th}=2,3$ in GRAND, using RLC (132,106), with $N_T=22$ and $N_R=138$, and 64-QAM. The corresponding PCH lower bounds are also plotted.}
   \label{fig:6}
\end{figure*}

The performance and the decoding complexity of the considered system were evaluated through numerical simulations. The number of codewords per MIMO channel use, previously defined in Section~\ref{sec:model}, is here set to $N_c=1$. In this setup, each codeword of $n$ bits is transmitted in ``one shot'', using Gray-coded $M$-QAM over a mMIMO channel with
$N_T={n}/ {\log_2(M)}$ transmit antennas. The antenna-sorting takes the whole set of $g_i$, for $i=1,\cdots, N_T$, and sorts them.

Two constellations were considered: $16$-QAM and $64$-QAM. As the $M$-arity of the modulation grows, the number of transmit antennas $N_T$ is decreased to accommodate the same payload of $n$ bits. The number of antennas was set to $N_T=32$ for \mbox{$16$-QAM} with a RLC $(128, 103)$, and $N_T=22$ for \mbox{$64$-QAM} with a RLC $(132, 106)$, such that the code rate $R=k/n=0.8$ is kept constant. Three different values for $N_R$ were tested for each $M$-arity: $N_R=50$, $100$ and $200$ for $16$-QAM, and $N_R=38$, $69$ and $138$ for $64$-QAM. To make the comparison between $16$-QAM and $64$-QAM fair, the number of receive antennas $N_R$ in each configuration was chosen to yield similar load factors $N_R/N_T$, i.e., $50/32\approx 38/22$, $100/32\approx 69/22$ and $200/32\approx 138/22$.

The antenna sorting preprocessing can be used to improve the decoding speed of symbol-level GRAND but also of bit-level GRAND. After arranging the bit-stings $\mathcal{S}(\tilde{s}_i)$ in ascending order of likelihood, one can also apply the original bit-level GRAND using its default flipping order for each bit. However, this leads to sub-optimal performance, given that the probability of the strings of $\log_2(M)$ bits is being used rather than the probability of the individual bits. We refer to this decoding method as \textit{sorted-bit-level decoding}. While sub-optimal, this ordering performs a step toward optimal bit ordering, and therefore reduces decoding complexity in comparison to standard unsorted bit-level GRAND.

Fig.~\ref{fig:1}, Fig.~\ref{fig:2} and Fig.~\ref{fig:3} illustrate the performance and decoding complexity results for 16-QAM, and Fig.~\ref{fig:4}, Fig.~\ref{fig:5} and Fig.~\ref{fig:6} show the performance and decoding complexity results for 64-QAM. The block error rate (BLER) has been used to evaluate systems performance as a function of ${E_b}/{N_0}$, as it is commonly used in recent works evaluating GRAND. The decoding complexity has been expressed in terms of the expected number of membership tests needed at each ${E_b}/{N_0}$. All figures include the curves for uncoded transmission, bit-level GRAND decoding, sorted-bit-level decoding, symbol-level GRAND decoding, and sorted-symbol-level decoding. Each system configuration is assessed with two different thresholds for the number of bits in error in the error pattern, $w_{th}=2,3$. The figures also include performance and complexity results when using symbol-level GRAND for PCH. Recall that, in the ideal scenario of PCH, antenna sorting prior to GRAND has no impact on the overall decoding complexity because all streams experience the same SNR after ZF detection, as seen in \eqref{eq:CH}.

As expected, the BLER performance greatly improves as one tests error patterns with larger weight, but this is achieved at the cost of a considerably larger number of membership tests. When considering a given $w_{th}$ threshold for the number of errors in a error pattern, an upper bound for the number of membership tests in the case of bit-level GRAND is
\begin{equation}
    UB= \sum_{t=0}^{w_{th}} \binom{n}{t} = 1 + \binom{n}{1} + \binom{n}{2} + \dots + \binom{n}{w_{th}},
\end{equation}
which is plotted in the figures showing the complexity results. Note that the first term in $UB$ accounts for the initial query that always needs to be carried out.
The results show that the average number of membership tests is much lower than the upper bound for $M=16$ with $N_R=200$ and $M=64$ with $N_R=138$. Nevertheless, when the noise is too large, the decoding complexity can get close to the upper bound due to the sheer number of erroneous symbols. As one would expect, when the noise vanishes, the average number of membership tests always tends to be one; in that case, all the received words are valid codewords and the only membership test performed is to check - and confirm - that the error pattern is $\mathbf{\hat{e}}_\text{b}=\mathbf{0}$.

One should note that the BLER performance results for bit-level GRAND, symbol-level GRAND, sorted-bit-level, and sorted-symbol-level for the analyzed range of $E_b/N_0$ are all the same (all four curves overlap). However, the complexity comparison illustrates that the sorted-antenna schemes and symbol-level GRAND remarkably outperform bit-level GRAND. The extra complexity reduction added by the sorting preprocessing becomes more significant when $N_R$ becomes smaller. This is due to a less strong channel hardening effect so that the equivalent SNR at each of the $N_T$ streams becomes more uneven. In that situation, antenna sorting becomes rather important. As seen in the complexity figures, there is a time saving of about 70\% when $M=16$ and up to 86\% in the system with 64-QAM. 

%The reason why the BLER gain effect only shows with $w_{th}=4$ is
%\begin{equation}
 %   \binom{n}{w_{th}} 2^{(n-k)}
%\end{equation}
%is only non negligible for $w_{th} \ge 4$. With $n=128$, we have 2.4223e-04 for $w_{th}=2$  and 0.0102 for $w_{th}=3$ and 0.3179 for  0.3179  $w_{th}=4$.
% that's all folks
%\twocolumn

\section{Conclusion}
\label{sec:conclusion}

This paper proposes a coded mMIMO transmission scheme for high-throughput, high-reliability, and low-latency, in accordance to the URLLC desiderata. This is accomplished using RLCs and ordered reliability symbol-level GRAND. Symbol-level GRAND is a variation of bit-level GRAND that considers the constellation structure of the adopted $M$-QAM scheme during the testing process of the error patterns that may explain the received sequence of bit strings.
The paper analyzes in detail symbol-level GRAND in the case of mMIMO with PCH. 
Then, when the channel conditions fall short from the ideal, it is shown that the ZF detector can provide a soft-metric for the reliability of each spatial stream, which in the proposed setup corresponds to a symbol reliability. 
The orthogonality defect of the mMIMO lattice is related to the variance of the reliability of the symbols. The disparity between the reliability of the symbols gets larger when $N_R$ decreases; in that case, the proposed antenna sorting can provide a significant reduction of the decoding complexity.
The results show that symbol-level GRAND provides much faster decoding times than the bit-level GRAND counterpart in the same mMIMO setup, throughout the SNR range of interest. The proposed antenna-sorting mechanism further speeds up the decoding process. The complexity reduction offered by symbol-level GRAND comes with a slight increase in memory usage. The extra sorting mechanism has linear complexity in respect to the number of spatial streams, $N_T$. It should be noted that the accentuated complexity reduction resulting from these techniques is accomplished without any observable performance degradation.

% if have a single appendix:
%\appendix[Proof of the Zonklar Equations]
% or
%\appendix  % for no appendix heading
% do not use \section anymore after \appendix, only \section*
% is possibly needed

% use appendices with more than one appendix
% then use \section to start each appendix
% you must declare a \section before using any
% \subsection or using \label (\appendices by itself
% starts a section numbered zero.)
%

%\appendices
%\section{Proof of the First Zonklar Equation}
% you can choose not to have a title for an appendix
% if you want by leaving the argument blank
%\section{}
%Appendix two text goes here.

\section*{Appendix}

\subsection{Probabilities of error types}

\begin{figure}[t]
    \centering
    \includegraphics[width=1 \columnwidth, clip=true, draft=false]{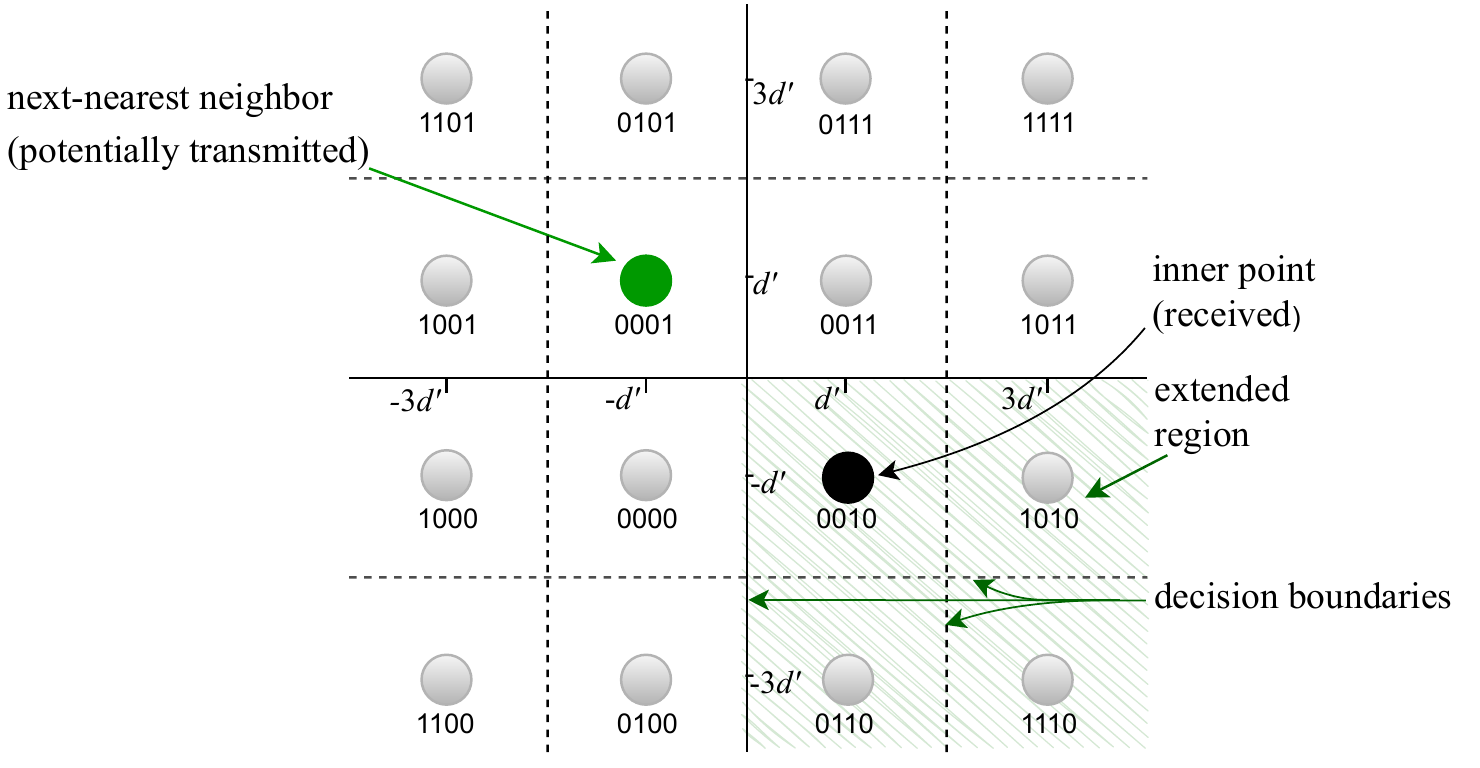}
    \caption{Example of calculating the probability that a type-$\mathcal{E}_2$ error string will be
added to 0010 and shift the received symbol to the region of
0001.}
    \label{fig:16-QAM_extended_compact}
\end{figure}

The derivation of the expressions in Table \ref{table_probabilities}, which compute the probability that a particular symbol was transmitted based on a received symbol, is here further detailed. These expressions were presented in \cite{Chatzigeorgiou_Monteiro_2023} but their derivation was not elaborated. In the example depicted in Fig. \ref{fig:16-QAM_extended_compact}, let $0010$ be the received symbol. At first, let us assume that $0010$ was actually transmitted. Note that $0010$ lies within a region defined by two horizontal and two vertical decision boundaries. The Euclidean distance between $0010$ and any of the four decision boundaries is $d'$. For the received symbol to match the transmitted symbol after hard detection, the noise should not cause the real and/or imaginary components of the transmitted symbol to cross the decision boundaries around the symbol. The probability that one of the components of the transmitted symbol will cross one of the decision boundaries and cause a decision error is given by $Q(d^{\prime})$, which represents the tail integral of a Gaussian function. Based on this, the following inferences can be made:
\begin{itemize}
\setlength\itemsep{-0.5mm}
  \item The probability that the real component of the transmitted symbol will cross the left-hand side vertical boundary is $Q(d^{\prime})$. Similarly, the probability that the real component of the transmitted symbol will cross the right-hand side vertical boundary is also $Q(d^{\prime})$.
  \item The probability that the real component of the transmitted symbol will cross the left-hand side vertical boundary \textit{or} the right-hand side vertical boundary is $Q(d^{\prime})+Q(d^{\prime})=2Q(d^{\prime})$.
  \item The probability that the real component of the transmitted symbol will cross \textit{neither} the left-hand side \textit{nor} the right-hand side vertical boundaries is $1-2Q(d^{\prime})$.
  \item The probability that the imaginary component of the transmitted symbol will cross \textit{neither} the top-side \textit{nor} the bottom-side horizontal boundaries is also $1-2Q(d^{\prime})$, due to symmetry.
  \item Therefore, the probability that the components of the transmitted symbol will \textit{not} cross any of the four boundaries and, thus, the symbol will be received correctly is  $(1-2Q(d^{\prime}))^2$.
\end{itemize}
The last derived expression $(1-2Q(d^{\prime}))^2$ has been assigned to $p_{\mathrm{i},0}$ in Table \ref{table_probabilities}, which represents the probability that an all-zero error string was added to a symbol mapped onto an inner point and, hence, did not alter its value.

% \begin{figure}
%     \centering
%     \includegraphics[width=1 \columnwidth, clip=true, draft=false]{Figures/16-QAM_extended_compact.pdf}
%     \caption{Caption}
%     \label{fig:enter-label}
% \end{figure}

The same methodology has been used to derive the nine probability expressions presented in Table \ref{table_probabilities}. For example, assume again that symbol $0010$ has been received, and we wish to calculate the probability that a type-$\mathcal{E}_2$ error string has been added to $0001$ such that it diagonally shifted it into the extended region highlighted in green in Fig. \ref{fig:16-QAM_extended_compact}. The calculation of this probability should take into account the following two conditions: \textit{i}) the components of the transmitted symbol will move beyond the bottom-side \textit{and} the right-hand side boundaries of $0001$; \textit{ii}) the components of the transmitted symbol will \textit{not} move beyond the bottom-side \textit{and} the right-hand side boundaries of $0010$.	
These two conditions perfectly define the region where $0010$ lies. Only decision boundaries closest to a potentially transmitted symbol were considered in order to simplify the probability expressions. Therefore, only the first condition is used, defining a larger region containing four points, including point $0010$. This region is part of the ``extended neighborhood 2'' of $0001$. As a result of this simplification, probability expressions that focus on type-$\mathcal{E}_1$ and type-$\mathcal{E}_2$ error strings in Table \ref{table_probabilities} are approximations.

% use section* for acknowledgment
\section*{Acknowledgments}
This work has been funded by Instituto de Telecomunica\c{c}\~{o}es and FCT/MCTES (Portugal) through national funds and when applicable co-funded EU funds under the projects UIDB/50008/2020. Sahar Allahkaram is funded by a merit scholarship from  Iscte - University Institute of Lisbon.
% Can use something like this to put references on a page
% by themselves when using endfloat and the captionsoff option.
\ifCLASSOPTIONcaptionsoff
  \newpage
\fi

\bibliographystyle{IEEEtran}
\bibliography{Main.bib}

%\vskip 0pt plus -1fil
\begin{IEEEbiography}
[{\includegraphics[width=1in,height=1.25in,clip,keepaspectratio]{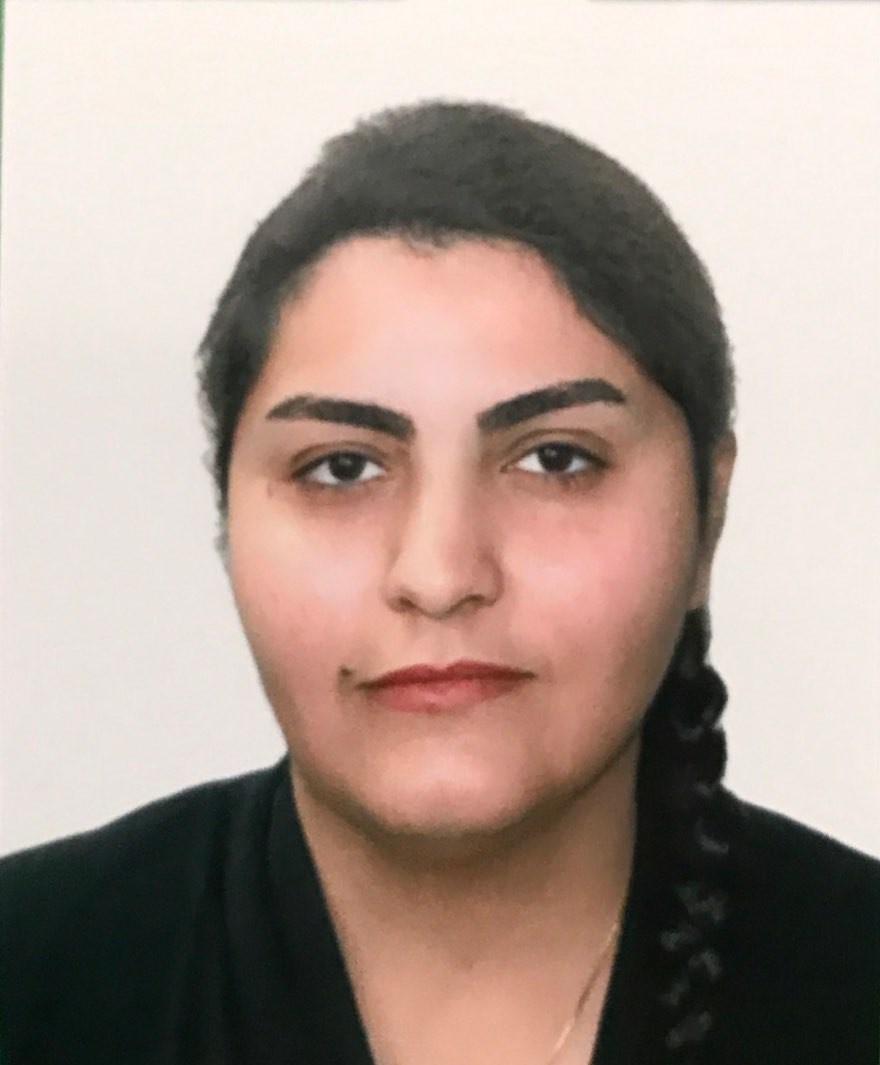}}]{Sahar Allahkaram} is a PhD student in the Dep. of Information Science and Technology at Iscte - University Institute of Lisbon, Portugal. She is working on Signal Processing and Coding Techniques for 6G Ultra-reliable Low-latency Wireless Machine-type Communications. She obtained her MSc in Aerospace Engineering from Sapienza University of Rome in 2020 and her BSc in Electronic Engineering from Azad University of Tehran in 2015.
\end{IEEEbiography}

%\vskip 0pt plus -1fil
\begin{IEEEbiography}
[{\includegraphics[width=1in,height=1.25in,clip,keepaspectratio]{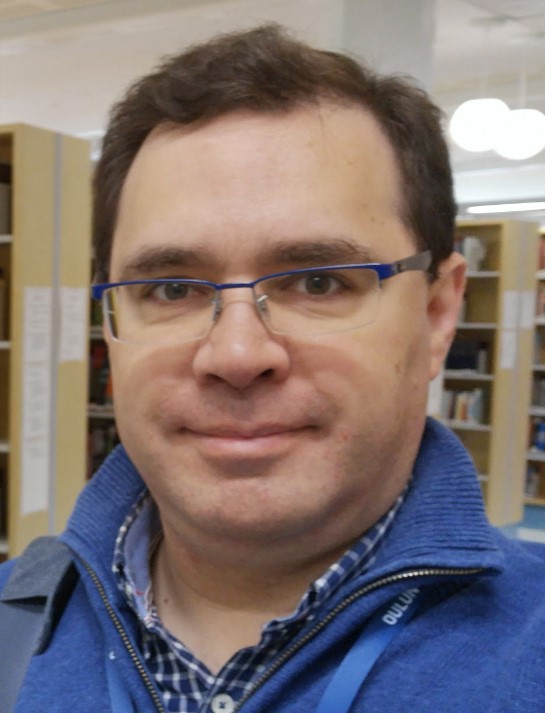}}]{Francisco A. Monteiro} (M'07) is Assistant Professor in the Dep. of Information Science and Technology at Iscte - University Institute of Lisbon, and a researcher at Instituto de Telecomunicações, Lisbon, Portugal. He holds a PhD from the University of Cambridge, UK, and the Licenciatura and MSc degrees in Electrical and Computer Engineering from IST, University of Lisbon, where he also became a Teaching Assistant. He held visiting research positions at the Universities of Toronto (Canada), Lancaster (UK), Oulu (Finland), and Pompeu Fabra (Barcelona, Spain). He has won two best paper prizes awards at IEEE conferences (2004 and 2007), a Young Engineer Prize (3rd place) from the Portuguese Engineers Institution (Ordem dos Engenheiros) in 2002, and for two years in a row was a recipient of Exemplary Reviewer Awards from the IEEE Wireless Communications Letters (in 2014 and in 2015). He co-edited the book ``MIMO Processing for 4G and Beyond: Fundamentals and Evolution'', published by CRC Press in 2014. In 2016 he was the Lead Guest Editor of a special issue on Network Coding of the EURASIP Journal on Advances in Signal Processing. He was a general chair of ISWCS 2018 - The 15th International Symposium on Wireless Communication Systems, an IEEE major conference in wireless communications.
\end{IEEEbiography}

%\vskip 0pt plus -1fil
\begin{IEEEbiography}
[{\includegraphics[width=1in,height=1.25in,clip,keepaspectratio]{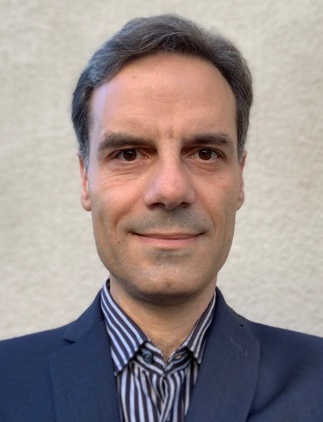}}]{Ioannis Chatzigeorgiou} (S’99–M’05–SM’15) is a Senior Lecturer at the School of Computing and Communications, Lancaster University, UK. He holds a Dipl.-Ing. degree in Electrical Engineering from Democritus University of Thrace, Greece, an MSc degree in Satellite Communication Engineering from the University
of Surrey, UK and a PhD degree from the University of Cambridge, UK. Prior to his appointment at Lancaster University, he worked at Marconi Communications and Inmarsat Ltd. He also held postdoctoral positions at the University of Cambridge and the Norwegian University of Science and Technology (NTNU) supported by the Engineering and Physical Sciences Research Council (EPSRC) and the European Research Consortium for Informatics and Mathematics (ERCIM), respectively. His research interests include communication theory with an emphasis on forward error correction, relay-aided communications and network coding.
\end{IEEEbiography}

\end{document}